\documentclass{article}

\usepackage{booktabs} 
\usepackage{tabularx} 
\usepackage{float}
\usepackage{threeparttable}
\usepackage{amsmath} 
\usepackage{arxiv}
\usepackage{bm}
\usepackage{amsmath,amsfonts,amssymb,bm}
\usepackage{mathptmx}
\usepackage{newtxtext}
\usepackage{newtxmath}
\usepackage{xcolor}
\usepackage[utf8]{inputenc} 
\usepackage[T1]{fontenc}    
\usepackage{hyperref}       
\usepackage{url}            
\usepackage{booktabs}       
\usepackage{amsfonts}       
\usepackage{nicefrac}       
\usepackage{microtype}      
\usepackage{cleveref}       
\usepackage{lipsum}         
\usepackage{graphicx}
\usepackage{natbib}
\usepackage{doi}
\usepackage{enumitem}

\usepackage{booktabs}
\usepackage{threeparttable}
\usepackage{tabularx}

\title{Deep Learning Atmospheric Models Reliably Simulate Out-of-Sample Land Heat and Cold Wave Frequencies}
\newif\ifuniqueAffiliation

\usepackage{authblk}

\newbox{\orcid}\sbox{\orcid}{\includegraphics[scale=0.06]{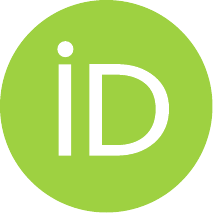}} 
\author[1]{%
	\href{https://orcid.org/0000-0001-5706-592X}{\usebox{\orcid}\hspace{2mm}Zilu Meng\thanks{Correspondence author: Zilu Meng, \texttt{zilumeng@uw.edu}}}%
}
\author[1]{%
	\href{https://orcid.org/0000-0001-8486-9739}{\usebox{\orcid}\hspace{2mm}Gregory J. Hakim}%
}
\author[2]{%
	\href{https://orcid.org/0000-0003-0053-9527}{\usebox{\orcid}\hspace{2mm} Wenchang Yang}%
}
\author[2, 3]{%
	\href{https://orcid.org/0000-0002-5085-224X}{\usebox{\orcid}\hspace{2mm} Gabriel A. Vecchi}%
}
\affil[1]{ 	Department of Atmospheric and Climate Science, University of Washington, Seattle, WA}
\affil[2]{  Department of Geosciences, Princeton University, Princeton, NJ}
\affil[3]{  High Meadows Environmental Institute, Princeton University, Princeton, NJ}

\renewcommand{\headeright}{Meng et al. 2025}
\renewcommand{\undertitle}{}
\renewcommand{\shorttitle}{DL models simulate Heatwave and Coldwave}

\hypersetup{
colorlinks,
linkcolor={red!50!black},
citecolor={blue!50!black},
urlcolor={blue!80!black}
}

\begin{document}
\maketitle

\begin{abstract}
Deep learning (DL)--based general circulation models (GCMs) are emerging as fast simulators, yet their ability to replicate extreme events outside their training range remains unknown. Here, we evaluate two such models---the hybrid Neural General Circulation Model (NGCM) and purely data-driven Deep Learning Earth System Model (DLESyM)---against a conventional high-resolution land--atmosphere model (HiRAM) in simulating land heatwaves and coldwaves. All models are forced with observed sea surface temperatures and sea ice over 1900--2020, focusing on the out-of-sample period (1900--1960). Both DL models generalize successfully to unseen climate conditions, broadly reproducing the frequency and spatial patterns of heatwave and coldwave events during 1900--1960 with skill comparable to HiRAM. An exception is over portions of North Asia and North America, where all models perform poorly during 1940--1960. Due to excessive temperature autocorrelation, DLESyM tends to overestimate heatwave and coldwave frequencies, whereas the physics--DL hybrid NGCM exhibits persistence more similar to HiRAM.
\end{abstract}

\section*{Key Points}
\begin{itemize}[label=•, leftmargin=*, nosep]
\item DL-based GCMs successfully simulate out-of-sample (1900–1960) heat and cold wave frequencies with skill comparable to a physical model
\item The degree of autocorrelation in the surface temperature field for the models roughly scales with the data-driven contribution to the model 
\item Locally poor performance over North Asia and North America suggests changes in the land-surface or weaker SST forcing 
\end{itemize}

\section*{Plain Language Summary}
Heatwaves and cold waves have major societal impacts, yet climate models often struggle to predict their occurrence. This study evaluates how well new deep learning (DL) atmospheric models simulate heatwaves and cold waves, especially when tested on early 20th-century conditions they have not encountered during training. We compare two DL-driven climate models with a traditional physics-based model, running them all with observed sea-surface temperatures from 1900 to 2020. We find that the DL models perform nearly as well as the traditional physical model in capturing heat and cold wave frequencies, and the probability of extreme events. Moreover, differences in model architecture influence frequency estimates by affecting the temporal persistence of surface temperature anomalies.

%
%
\section{Introduction} \label{intro}
Extreme multi-day near-surface temperature events--heatwaves and coldwaves, are essential components of climate and weather variability that have significant societal and ecological impacts \citep[e.g.,][]{legg2021ipcc,canton2021world,ncei_billion_dollar_2025}. Accurately simulating and predicting their frequency, intensity, and spatial distribution remains a key challenge. Despite advances in traditional physical general circulation models (GCMs), they still often exhibit limited skill in simulating these extremes \citep[e.g.,][]{domeisenPredictionProjectionHeatwaves2022, tebaldi2006going,orlowsky2012global,hirsch2021cmip6}. Because heatwaves and coldwaves are intrinsically uncommon and influenced by multiscale processes that include chaotic fluctuations in the Earth system, correctly understanding and predicting these extremes benefits from the ability to run many simulations, which in turn requires computationally efficient tools  \cite[e.g.,][]{ragone2018computation,gessner2021very,seneviratne2021weather}.

Recent advances in data-driven deep learning (DL) models offer an alternative approach to dynamically modeling climate variability. DL models trained on reanalysis data \citep[e.g.,][]{hersbachERA5GlobalReanalysis2020} have shown promising results in weather forecasting, with skill similar to operational models \citep[e.g.,][]{biAccurateMediumrangeGlobal2023,lam2023learning,kochkovNeuralGeneralCirculation2024a,chen2023fuxi}. Moreover, these models have also shown the ability to predict weather extremes such as the heatwaves, tornadoes, and hurricanes \citep[e.g.,][]{biAccurateMediumrangeGlobal2023,lam2023learning,chen2023fuxi,vonich2024predictability,hua2025performance, chienModulationTropicalCyclogenesis2025, hua2023tornadic}. This success motivates extending the use of these models from weather forecasting to climate simulation, including fully data-driven models \citep[e.g.,][]{watt-meyerACE2AccuratelyLearning2024,cresswell-clayDeepLearningEarth2024} and physics-DL hybrid models \citep{kochkovNeuralGeneralCirculation2024a} to produce stable climate simulations with realistic emergent phenomena, such as the frequency and tracks of tropical cyclones, and multi-year to multi-decadal land and atmospheric temperature trends when given prescribed boundary conditions (e.g., sea surface temperature (SST) and sea ice concentration (SIC)). These studies suggest, once trained, DL-driven models can, under certain setups, reproduce important aspects of climate variability at a fraction of the computational cost of conventional GCMs.

While the models appear capable of reproducing internal atmospheric variability during the time period on which they are trained, it is less clear whether these findings hold for out-of-sample time periods \citep{ullrich2024recommendations}. Most existing studies validate model performance under the climate conditions with abundant training data existing \citep[e.g.,][]{van2025reanalysis,watt-meyerACE2AccuratelyLearning2024,kochkovNeuralGeneralCirculation2024a,cresswell-clayDeepLearningEarth2024, chienModulationTropicalCyclogenesis2025}, which tends to overlap or be similar to the training set. For example, \cite{kochkovNeuralGeneralCirculation2024a} evaluated global mean temperature from 1980 to 2020 while the model is trained on ERA5 \citep{hersbachERA5GlobalReanalysis2020} from 1979 to 2017. A large overlap between the training and verification periods limits assessments of model generalization. Moreover, it is uncertain to what extent these models truly capture the underlying physics versus implicitly learning the particular trends and variability of the training era. To fill this critical gap, we present an evaluation of DL-based GCMs and compare their performance to a conventional physical GCM verified against reanalysis data from 1900 to 2014. All models are forced with the same observed sea surface temperature (SST) and sea ice concentration (SIC) data, following the AMIP (Atmospheric Model Intercomparison Project) protocol \citep[e.g.,][]{gates1999overview}, enabling a direct comparison. Here, we focus on extreme land-based temperature events, specifically heatwaves and coldwaves. 

Another important question concerns the influence of DL model architecture on their ability to simulate extreme events. Unlike traditional physical models, which are grounded in explicit physical theories and parameterizations that can be formally expressed mathematically, DL-based models can adopt a wide range of architectures shaped by engineering decisions, such as training efficiency, inference cost, and forecast accuracy. DL models used for weather and climate prediction span several major architecture families, including multilayer perceptrons (MLPs), convolutional neural networks (CNNs, e.g., U-Nets), recurrent neural networks (RNN), transformers, and physics–DL hybrids \citep[e.g.,][]{goodfellow2016deep}. These model families differ in their built-in assumptions about data (inductive biases), their learning behavior during training (training dynamics), and the balance they strike between computational efficiency and physical interpretability \citep{goodfellow2016deep,reichsteinDeepLearningProcess2019a,wuDataDrivenWeatherForecasting2024}. As a result, a key question arises: how does model architecture affect the ability of models to simulate climate variability and extremes? In this study, we compare three models with differing levels of physical and data-driven components: Deep-Learning Earth System Model (DLESyM), a fully deep learning–based GCM \citep{cresswell-clayDeepLearningEarth2024}; Neural-GCM (NGCM) a physics–DL hybrid model that blends numerical dynamics with learned subgrid processes \citep{kochkovNeuralGeneralCirculation2024a}; and High Resolution Atmospheric Model (HiRAM), a conventional high-resolution physical model \citep{zhaoSimulationsGlobalHurricane2009}. These models span a continuum from purely data-driven (DLESyM) to fully physics-based (HiRAM), allowing us to assess how the degree of explict physical constraints in the model architecture influences performance in simulating climate extremes.


Therefore, in this study, we aim to address the following research questions:
\begin{enumerate}
\item How well can the frequency of heatwaves and coldwaves be simulated by prescribing only observed sea surface temperature (SST) and sea ice concentration (SIC)?
\item How well do DL-based GCMs reproduce heatwave and coldwave frequencies during the out-of-sample period 1900–1960 when compared with reanalysis datasets and a traditional physical GCM?
\item How do purely physical, hybrid, and purely data-driven models differ in simulating heatwaves and coldwaves?
\end{enumerate}

The remainder of the paper aims to answer these questions with organization as follows. Section~\ref{sec:models} describes the models, datasets, and methods used in this study. Section~\ref{sec:results} presents the results, including heatwave and coldwave frequency time series, spatial patterns and correlations with reanalysis data. Finally, Section~\ref{sec:conclusion} discusses the implications of our findings and provides conclusions.

\section{Models, Data, and Methods} \label{sec:models}

\subsection{Models}

In this subsection, we describe the models used in this study. Specifically, we utilize three types of models: a purely data-driven DL-based GCM, DL\textit{ESy}M \citep{cresswell-clayDeepLearningEarth2024}; a physics–DL hybrid model, NGCM \citep{kochkovNeuralGeneralCirculation2024a}; and a traditional physics-based atmospheric GCM, HiRAM \citep{zhaoSimulationsGlobalHurricane2009}. Although another purely data-driven DL-based GCM, ACE2-ERA5 \citep{watt-meyerACE2AccuratelyLearning2024}, was tested in this research, we do not include it in our full evaluation because its training period (1940--1995 and 2011--2019) substantially overlaps with our designated out-of-sample validation period (1900–1960), which could compromise the assessment of model generalization. A brief summary of the GCMs used in our paper appears in Supplementary Table \ref{tab1}. Additionally, to analyze the interannual temperature variability of the AMIP experiments, we introduce a simple linear model.

\subsubsection{Deep Learning Earth SYstem Model (DL\textit{ESy}M)}

The DL\textit{ESy}M is a purely data-driven DL-based GCM \citep{cresswell-clayDeepLearningEarth2024}. The model uses a U-net architecture \citep{ronneberger2015u} that takes as input the current state $\mathbf{x}_t$ and 6-hour-earlier state $\mathbf{x}_{t-6hr}$ to predict next 6-hour state $\mathbf{x}_{t+6hr}$:
\begin{equation}
    \mathbf{x}_{t+6hr} = \text{DLWP} (\mathbf{x}_t, \mathbf{x}_{t-6hr}),
\end{equation}
\noindent which may be recursively extended indefinitely. The DL\textit{ESy}M is a ocean-atmosphere coupled model, but only the atmosphere component is predicted for this study as discussed below. The atmospheric model is trained using ERA5 reanalysis data \citep{hersbachERA5GlobalReanalysis2020}, while outgoing longwave radiation (OLR) is trained on observational data from the International Satellite Cloud Climatology Project (ISCCP) \citep{rossow2004international} for the period 1983––2016.

\subsubsection{Neural General Circulation Model (NGCM)}

The Neural General Circulation Model (NGCM) is a physics--DL hybrid model, which makes forecasts by combining physical equations with a deep learning parameterization, as represented by \citep{kochkovNeuralGeneralCirculation2024a}:
\begin{equation}
    \frac{\partial \mathbf{x}}{\partial t} = \Phi(\mathbf{x}) + \Psi(\mathbf{x}).
\end{equation}
\noindent Here $\mathbf{x}$ denotes the state vector (e.g., vorticity and divergence), $\Phi$ represents the tendency from physical conservation laws (e.g., vorticity and divergence tendency equations), and $\Psi$ is a deep learning model that has been trained on the residual between the physical equations and the ERA5 reanalysis dataset \citep{hersbachERA5GlobalReanalysis2020} during 1979--2017. In this study, we use the NGCM-2.8° deterministic model, which shows stable climate simulations as discussed in \cite{kochkovNeuralGeneralCirculation2024a}. To maintain model stability over multi-century simulations, the global mean surface pressure is fixed to a constant value following the approach in \citet{neuralgcm2024}.



\subsubsection{High-Resolution Atmospheric Model (HiRAM)}

The Geophysical Fluid Dynamics Laboratory (GFDL) High-Resolution Atmospheric Model (HiRAM) \citep{zhaoSimulationsGlobalHurricane2009} is a traditional  physical atmospheric GCM developed at Geophysical Fluid Dynamics Laboratory (GFDL). While the details are complex, HiRAM can be summarized as:
\begin{equation}
    \frac{\partial \mathbf{x}}{\partial t} = \Phi(\mathbf{x}),
\end{equation}
\noindent where $\Phi$ represents the physical governing equations along with parameterization schemes. At the horizontal grid spacing considered here (around 50km), which is finer than that of typical climate models, HiRAM is capable of simulating the statistics of small-scale features such as tropical cyclones, as demonstrated by \citet{zhaoSimulationsGlobalHurricane2009}, \citet{harris2016high}, \citet{Chan2021}, and \citet{Yang2021}.

\subsubsection{Linear Statistical Model} 

To further understand the interannual temperature variability of the AMIP experiments, we train a simple linear model over the period 1980–2020, corresponding to the main training period of the deep learning models. We first apply Empirical Orthogonal Function (EOF) truncation using SACPY \citep[e.g.,][]{meng2023sacpy} on the HadISST dataset \citep{rayner2003global} from 80°N to 80°S to derive the first three spatial patterns and their associated principal components (PCs) time series. 
We then perform a multivariate linear regression to relate the annual mean regional temperature \(T(t)\)to these PCs:
\begin{equation}
T(t) = \sum_{i=1}^{3} k_i\,x_i(t) + b,
\end{equation}
where \(k_i\) denotes the regression slope for the \(i\)th EOF, \(x_i(t)\) is the corresponding PC, \(t\) is the time, and \(b\) is the intercept. The EOF spatial patterns and PCs are presented in Figure \ref{sfig:eof}. Additional tests using EOFs beyond the first three (3––10) yield consistent results, indicating that the number of EOFs selected does not affect our conclusions.

\subsection{Datasets}


20th Century Reanalysis Version 3 (20CRv3) \citep{slivinski2021evaluation} is a reanalysis dataset from 1836 to 2015 that we use to verify statistics of temperature extremes. It derives from assimilating surface pressure observations over ocean regions using an ensemble Kalman filter subject to prescribed sea surface temperature (SST) and sea ice concentration (SIC) boundary conditions. Here we use all 80 ensemble members to calculate the heatwave and coldwave frequencies. We note that ERA20C \citep{poliERA20CAtmosphericReanalysis2016} is similar to 20CRv3, but since it only provides the ensemble mean, is insufficient for capturing event frequency distributions using the method discussed below. Additionally, ERA5 \citep{hersbachERA5GlobalReanalysis2020}, covering 1940 to the present and incorporating a wide range of satellite and in situ observations, is used to verify our results and to compare with 20CRv3 ensemble outputs for extreme event analysis from 1960 to 2010. To further assess the reliability of 20CRv3 in capturing long-term temperature variability, we use the Berkeley Earth (BE) dataset \citep{rohde2020berkeley}, which provides land and ocean surface temperatures from 1850 to the present. However, BE is not used directly for calculating heatwave and coldwave frequencies due to missing data between 1900 and 1960, except for selected locations where a direct comparison is enabled by nearly complete observation time series. As shown in Figure~\ref{sfig:temp_trend}, the long-term temperature trends over land in regions such as Central Asia and North America differ between 20CR and BE. To further validate our experimental results, we construct a new dataset—referred to as 20CR-BE,by combining the annual mean temperatures from BE with the daily anomalies from 20CRv3 (i.e., deviations from the 20CRv3 annual mean).

\subsection{Experiments Design}
We conduct experiments following an Atmospheric Model Intercomparison Project (AMIP) protocol \citep[e.g.,][]{gates1999overview,eyring2016overview,meng2024pacific} using NGCM, DL\textit{ESy}M, ACE2-ERA5, and HiRAM. These experiments are driven by SST and SIC boundary conditions from 1900 to 2020, based on the HadISST dataset \citep{rayner2003global}. The chosen period allows for evaluation of model generalization capability: NGCM and DL\textit{ESy}M are trained on ERA5 reanalysis data from 1980 to 2020, enabling 1900–1960 to serve as an out-of-sample period, and 1961–2010 as an in-sample period. It is worth noting that the global mean temperature during 1961–2010 is approximately 0.5°C higher than that during 1900–1960, providing a modest test of model resilience to changing climate.

We note that DL\textit{ESy}M only uses SST boundary condition, whereas NGCM and HiRAM are forced with both SST and SIC. Additionally, for radiative forcings except volcanic forcings, HiRAM follows the CMIP5 ``historical scenario" through 2005 and RCP4.5 thereafter \citep{taylor2012overview}. For volcanic forcings, due to significant change from CMIP5 to CMIP6 \citep{Yang2019, Jacobson2020},  HiRAM  follow the CMIP6 historical scenario through 2014 \citep{o2016scenario} and ssp245 thereafter. Initial conditions are randomly sampled from ERA5 reanalysis data between 1980 and 2020 for NGCM and DL\textit{ESy}M. We conduct 100 ensemble simulations for both NGCM and DL\textit{ESy}M, and 5 ensemble simulations for HiRAM. Differences in ensemble size are accounted for using bootstrap sampling in frequency comparisons as described below.

\subsection{Heatwave and Coldwave Frequency Calculation}

Heatwave frequency is calculated following the methods from \citet{hirsch2021cmip6} and \citet{perkinsIncreasingFrequencyIntensity2012}. To define heatwave days, we first compute the daily temperature exceedance for each grid point, defined as the difference between the daily mean 2-meter air temperature (calculated as the average of all time-step temperatures for that day) and the 90th percentile of daily mean 2-meter temperatures for the corresponding calendar day. For each ensemble member and each calendar day, the 90th percentile is calculated using a 15-day window centered on the day, across all available years, resulting in approximately 900 (15 $\times$ 60 years) daily values per grid point for percentile estimation. A consecutive three-day threshold is then applied: temperature exceedance must remain positive for at least three consecutive days to qualify as a heatwave event. Coldwave days are defined similarly, except based on the 10th percentile, and exceedance must be negative for at least three consecutive days. To evaluate model generalization outside the training period, we calculate the heatwave and cold wave indices separately for two periods: 1900–1960 and 1961–2010, meaning different 90th percentile of daily mean 2-meter temperatures for these two periods. 

We do not apply any seasonal filtering, as heatwaves and coldwaves can occur in any season, which are also called heat spells and cold spells \citep{perkinsIncreasingFrequencyIntensity2012}. For NGCM, which does not predict 2-meter air temperature directly, we use 1000 hPa temperature as a proxy, as we find that 1000 hPa temperature and 2-meter temperature are highly correlated and yield consistent results (Figure~\ref{sfig:era5t1000t2}). It is important to note that heatwave and coldwave frequencies depend on the three-day persistence criteria. As a result, these frequencies are strongly influenced by the persistence of temperature exceedance, which is reflected in the temperature autocorrelation. In particular, since the ERA-20C dataset provides only ensemble mean fields, which exhibit substantially higher autocorrelation compared to individual ensemble members (Figure~\ref{sfig:currentAuto}), this results an overestimation of heatwave and cold wave frequencies in ERA-20C relative to datasets that provide individual ensemble members. Consequently, we do not use ERA20C for frequency validation.

\section{Results} \label{sec:results}

\begin{figure}
    \centering
    \includegraphics[width=1.1\linewidth]{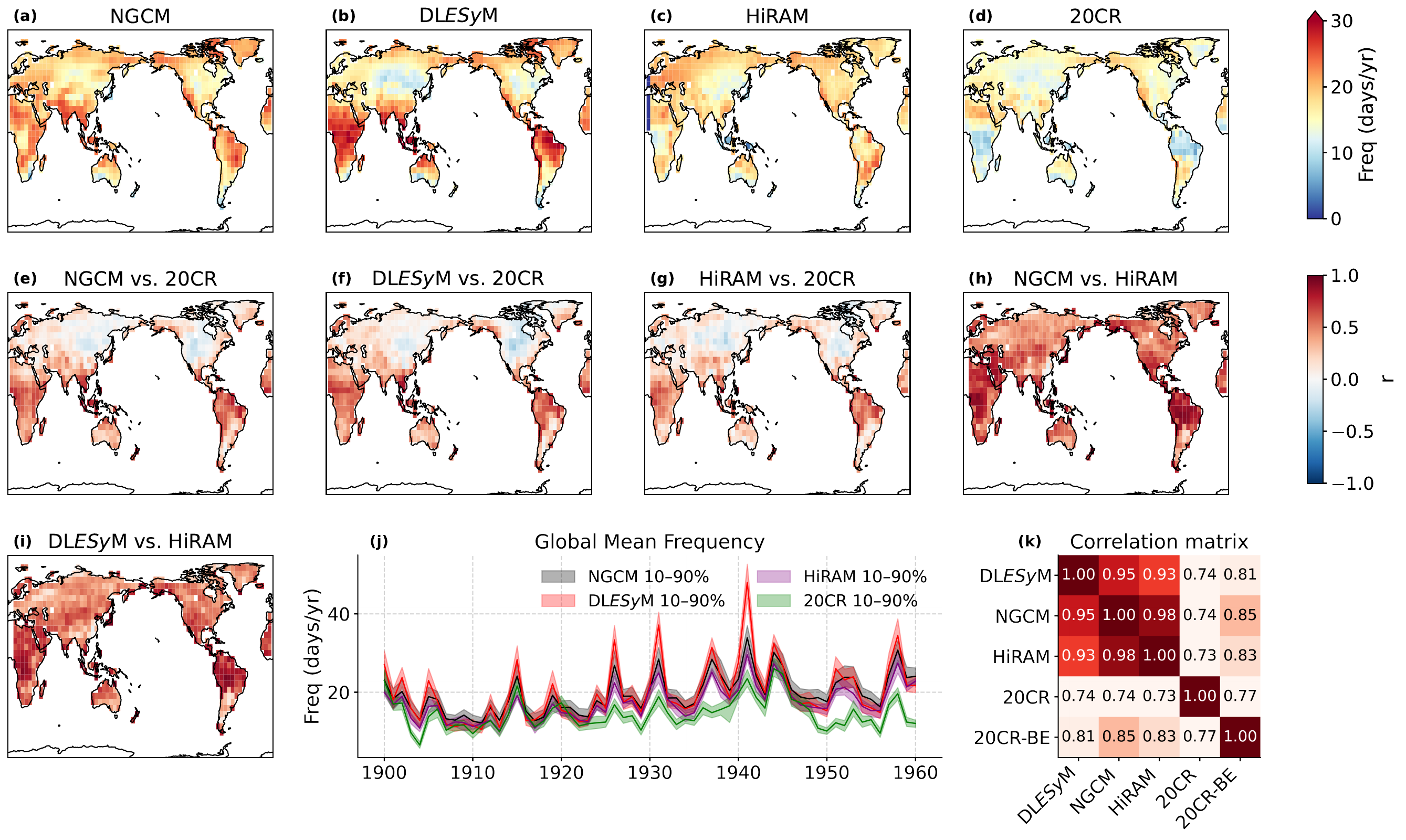}
    \caption{\textbf{Heatwave Frequency.} (a--d) Time-average heatwave frequency from 1900 to 1960 in NGCM, DL\textit{ESy}M, HiRAM, and 20CRv3. (e--g) Correlation of annual mean heatwave frequency with 20CRv3. (h--i) Correlation of annual mean heatwave frequency in HiRAM with DL\textit{ESy}M and NGCM. (j) Global mean of annual mean heatwave frequency as a function of time from 1900 to 1960. (k) Correlation matrix of global-mean annual heatwave frequency between each model, 20CRv3 and 20CR-BE.}
    \label{fig:heatwave}
\end{figure}

\begin{figure}
    \centering
    \includegraphics[width=1.1\linewidth]{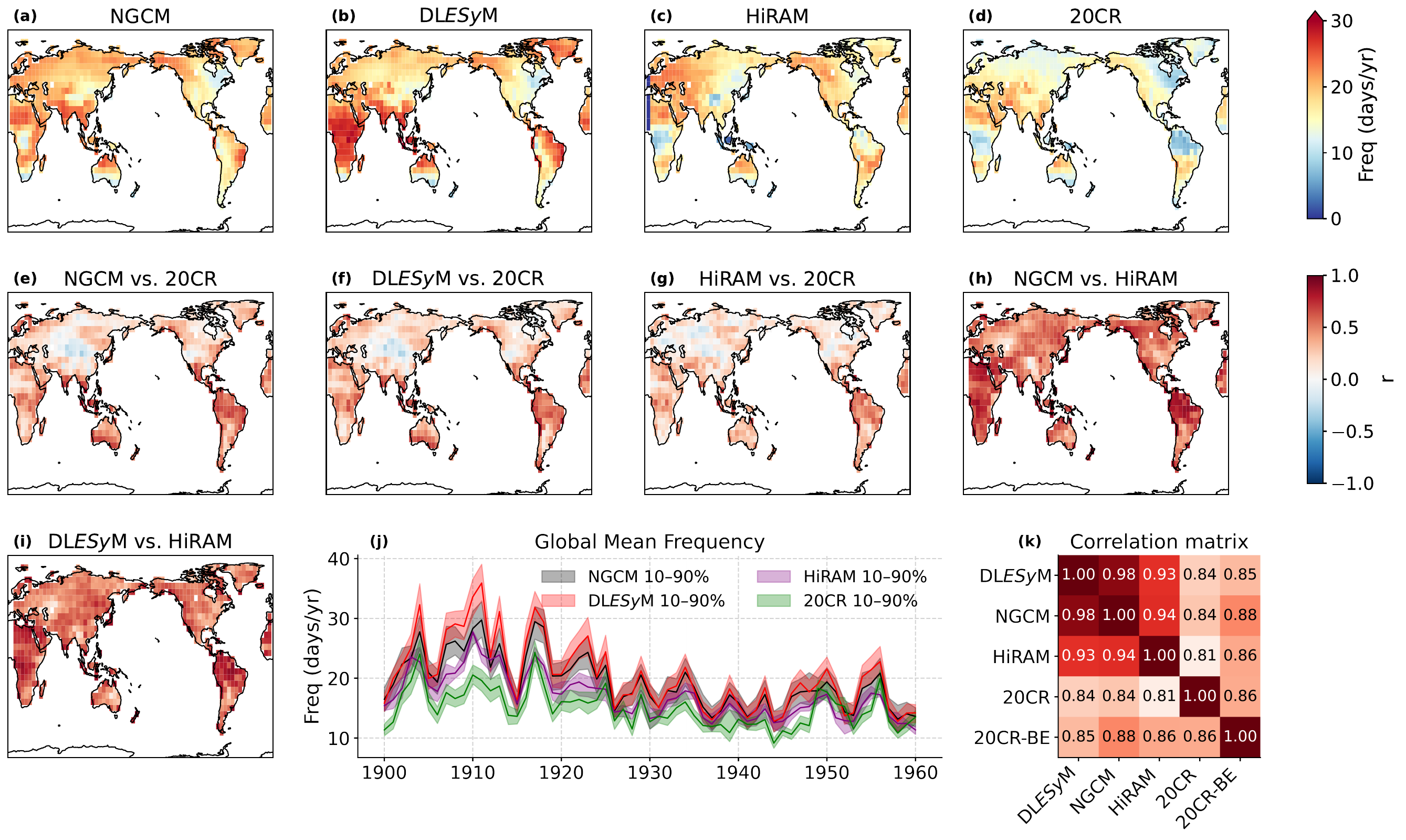}
    \caption{\textbf{Coldwave Frequency.} As in Figure~\ref{fig:heatwave}, but for coldwaves.}
    \label{fig:coldwave}
\end{figure}

\begin{figure}
    \centering
    \includegraphics[width=\linewidth]{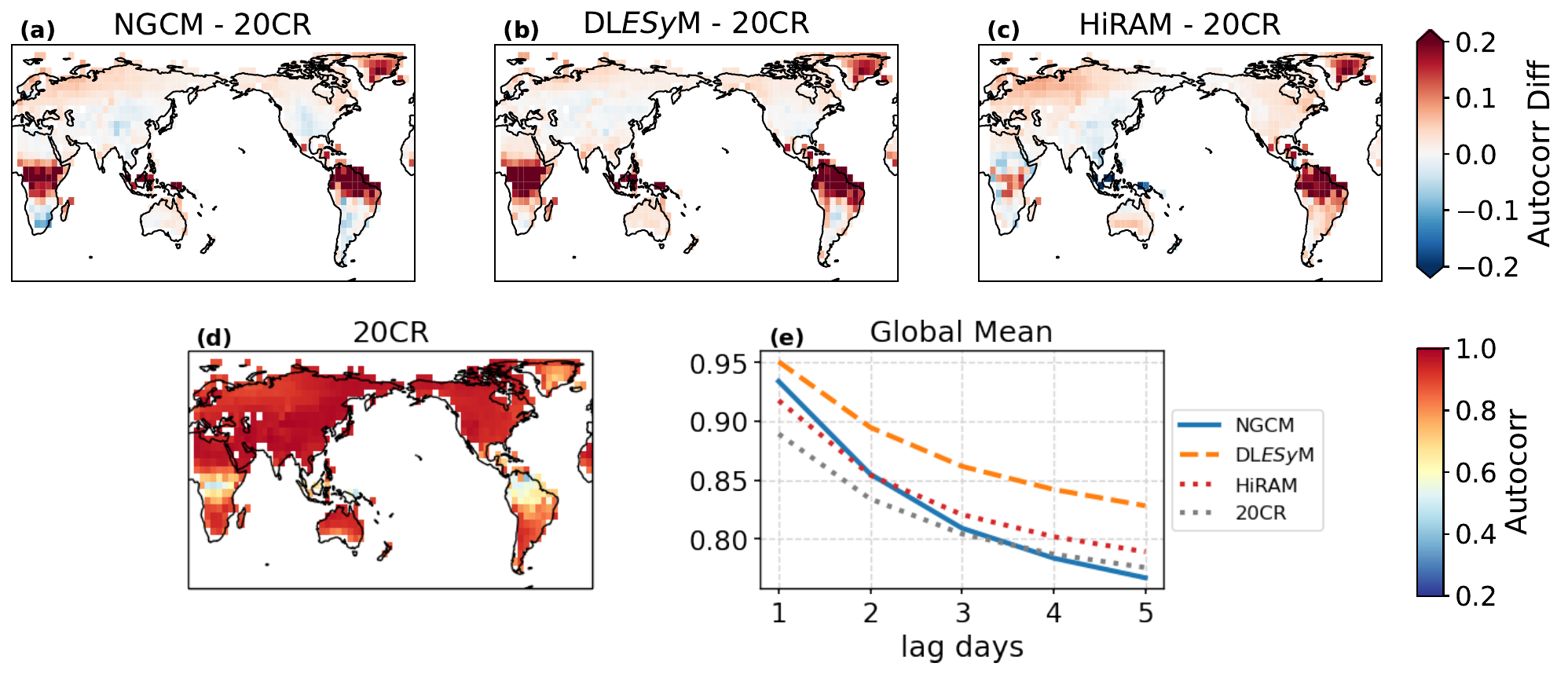}
    \caption{\textbf{Autocorrelation of Daily Mean Temperature (1900--1960).} (a--c) Differences in 1-day lag autocorrelation: (a) NGCM minus 20CRv3, (b) DL\textit{ESy}M minus 20CRv3, and (c) HiRAM minus 20CRv3. (d) 1-day lag autocorrelation for 20CRv3. (e) Global-mean land temperature autocorrelation as a function of lag time for NGCM (blue, solid), DL\textit{ESy}M (yellow, dashed), HiRAM (red, dotted), and 20CRv3 (gray, dotted).}
    \label{fig:autocorr}
\end{figure}

As shown in Figures~\ref{fig:heatwave} and \ref{fig:coldwave}, during the out-of-training period (1900––1960), the model-simulated spatial patterns of heatwave and coldwave frequency are broadly similar to those in 20CRv3, but with a high bias, particularly in tropical South America, central Africa, and Greenland. HiRAM exhibits smaller biases than NGCM and DL\textit{ESy}M, as expected from a physically based model, yet still overestimates frequency relative to 20CRv3. In the more recent period (1960–2010), HiRAM's results align more closely with ERA5 (Figures~\ref{sfig:currentHW} and \ref{sfig:currentCW}). There are notable differences between 20CRv3 and ERA5, particularly over Equatorial Africa, Greenland and Northern South Africa. Model overestimation of heatwave and coldwave frequency is primarily associated with higher autocorrelation of daily mean surface temperature (Figure~\ref{fig:autocorr}) as there is no large difference (mostly less than 2\%) in the 90th percentile and 10th percentile threshold (Figure~\ref{sfig:hwthres} and \ref{sfig:cwthres}). Positive autocorrelation differences relative to 20CRv3 are spatially consistent with the frequency over-estimation in North Asia, Greenland, and central Africa. Higher autocorrelation implies greater persistence of temperature anomalies, which increases the likelihood of exceeding heatwave or coldwave consecutive days thresholds. Autocorrelation differences can be partly attributed to the AMIP SST protocol, which uses interpolated monthly SST rather than actual daily SST. As shown in Figure~\ref{sfig:ERA5-AMIP}, this temporal interpolation increases autocorrelation over tropical oceans, increasing autocorrelation over nearby tropical land regions, but not globally. Despite the overestimation, the global-mean heatwave and coldwave frequencies during 1900–1960 are well correlated with 20CRv3, with correlation coefficients across all models of 0.73--0.74 for heatwaves and 0.81--0.84 for coldwaves. Replacing the annual variability of 20CRv3 with that of BE results in a small increase in correlation for all models (Figures~\ref{fig:heatwave}k and \ref{fig:coldwave}k).

\begin{figure}
    \centering
    \includegraphics[width=\linewidth]{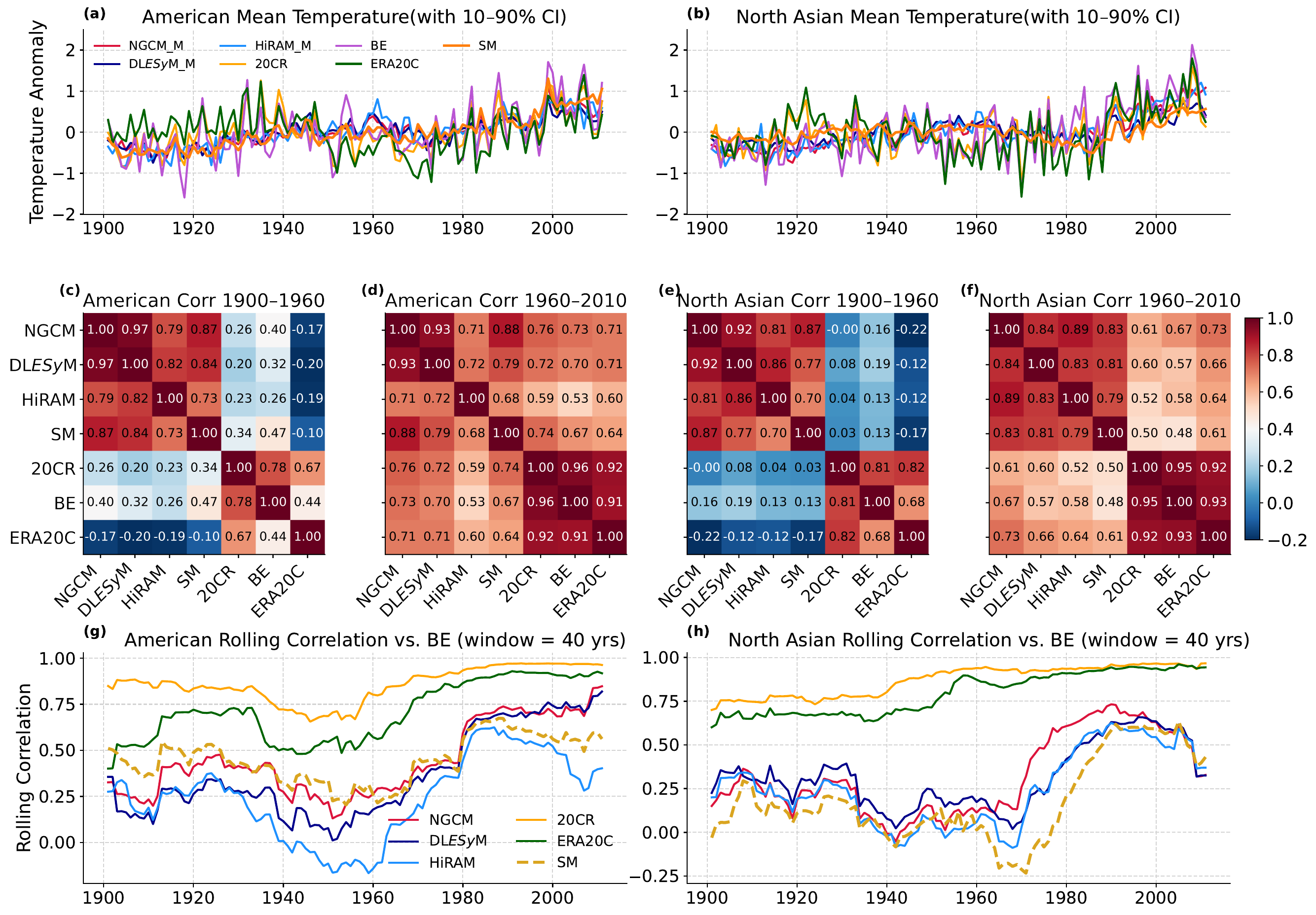} 
    \caption{\textbf{Annual Temperature over America (Left: 125°W--71°W, 20°N--58°N) and North Asia (Right: 70°E--160°E, 40°N--73°N).} (a–b) Annual ensemble mean temperature over America (a) and North Asia (b). (c–f) Correlation matrices between all models, 20CR, Berkeley Earth (BE), and ERA20C during 1900--1960 (c, e) and 1960--2010 (d, f). (g–h) 40-year rolling correlations with Berkeley Earth. The spatial extent of the two regions is shown in Figure~\ref{sfig:region}.}
    \label{fig:annualmean}
\end{figure}

All models show positive correlations with 20CRv3 across most regions, with particularly high correlations in the tropics as SST signal is strong in the tropical land (Figures~\ref{fig:heatwave}e–g and \ref{fig:coldwave}e–g). Exceptions include North Asia and parts of the Americas, where the correlations are relatively weaker. When comparing NGCM and DL\textit{ESy}M with HiRAM (Figures~\ref{fig:heatwave}h–i and \ref{fig:coldwave}h–i), correlations are consistently positive globally, indicating strong agreement between the three models. During the in-sample period (1961–2010), the spatial correlation with 20CRv3 improves, particularly over North Asia and the Americas, in both the physical and DL-based models (Figures~\ref{sfig:currentHW} and \ref{sfig:currentCW}). We also examined one simulation for the ACE2-ERA5 model and found that its performance is broadly similar to that of individual members of DL\textit{ESy}M, both during the in-sample period (1980–2020) and the out-of-sample period (1900–1960) (figure not shown).  We attribute this improvement to better representation of interannual variability in the forcing SST during the later period, while the AMIP experiments show limited skill in reproducing such variability prior to 1960 (Figure~\ref{fig:annualmean}). This enhancement in skill may also be partially due to a higher signal-to-noise ratio in recent decades, which facilitates the emergence of predictable patterns \citep{jia2016roles}. 

From 1900 to 1960, the annual temperature correlations between the three models and the three observational datasets (including reanalyses) are generally weaker compared to the 1960–2010 period. Rolling correlation results (Figures~\ref{fig:annualmean}e and \ref{fig:annualmean}h) further highlight periods of reduced correlation from the 1940s to the 1960s across all models. Using the simple linear model (SM) defined in section \ref{sec:models}, we find that the decadal variability in correlation skill can be reasonably captured by the leading three modes of SST variability (recall SM is trained during 1961--2010). This suggests that the frequency variability may be related to nonstationary teleconnections to SST forcing, rather than radiative forcing from aerosols, which are absent from all models but HiRAM. Additionally,  land-surface changes, such as aridity over North American during the dust bowl period of the 1930s, are not captured in AMIP experiments \citep{donatExtraordinaryHeat1930s2016}. Another potentially contributing factor may be the lower quality of SST observations in the early 20th century \citep[e.g.,][]{sippelEarlytwentiethcenturyColdBias2024, chanSystematicDifferencesBucket2019}. Another contributing factor may be the relatively low signal-to-noise ratio in SST forcing during the mid-20th century, when global SST variability was weaker, reducing the strength of the boundary-driven signal. For coldwaves, the low correlation over the Tibetan Plateau may be related to limited training data in this region, the relatively low reliability of reanalysis products over complex terrain \citep{baoImpactObservationData2020}, and for NGCM to the interpolation of temperature to 1000 hPa. 

HiRAM exhibits reduced skill in capturing interannual temperature variability over America during 1940--1960, with correlation values lower than the 100-member ensemble means of both NGCM and DL\textit{ESy}M (Figure~\ref{fig:annualmean}g). This lower performance can be partly attributed to HiRAM's limited ensemble size (5 members). As shown in Figure~\ref{sfig:bstest}, HiRAM's correlation falls within the 5th–95th percentile range of the NGCM 5-member bootstrap distribution.

By comparing results across the three models, we can draw preliminary conclusions about how the degree to which physical constraints may influence extreme temperature event simulation. The fully DL–based DL\textit{ESy}M shows the highest temperature autocorrelation, followed by the hybrid physics–DL NGCM, and then the fully physics-based HiRAM. This difference in autocorrelation results in overestimation of heatwave and coldwave frequencies in DL\textit{ESy}M relative to the other two models. For comparison, another fully DL-based model, ACE2-ERA5, also shows higher autocorrelation than NGCM and HiRAM, but lower than DL\textit{ESy}M (Figure~\ref{sfig:currentAuto}).

To assess whether out-of-sample skill primarily reflects recurrence of boundary forcing, we perform an analog-year analysis. For each out-of-sample year (1900--1960), we compute area-weighted spatial correlations between the annual-mean SST anomaly field and those from every training year (1979--2017); the in-sample year with the maximum correlation is designated the analog. We then compare, for each model, the simulated heatwave and coldwave frequency patterns between the out-of-sample year and its analog, and summarize the association between SST-pattern similarity and extremes-pattern similarity across all out-of-sample years (Figure~\ref{sfig:corrsst}). For heatwaves, the SST--extremes linkage is weak: $r=0.19$ for DL\textit{ESy}M, $r=0.30$ for NGCM, and $r=0.40$ for HiRAM. Coldwaves show somewhat stronger relationships, $r=0.63$  for DL\textit{ESy}M, $r=0.66$ for NGCM, and $r=0.50$ for HiRAM, yet the DL models (DL\textit{ESy}M, NGCM) exhibit correlations of similar magnitude to the physical model (HiRAM), indicating that recurrence of SST patterns alone is insufficient to explain out-of-sample performance. 

\begin{figure}
    \centering
    \includegraphics[width=1\linewidth]{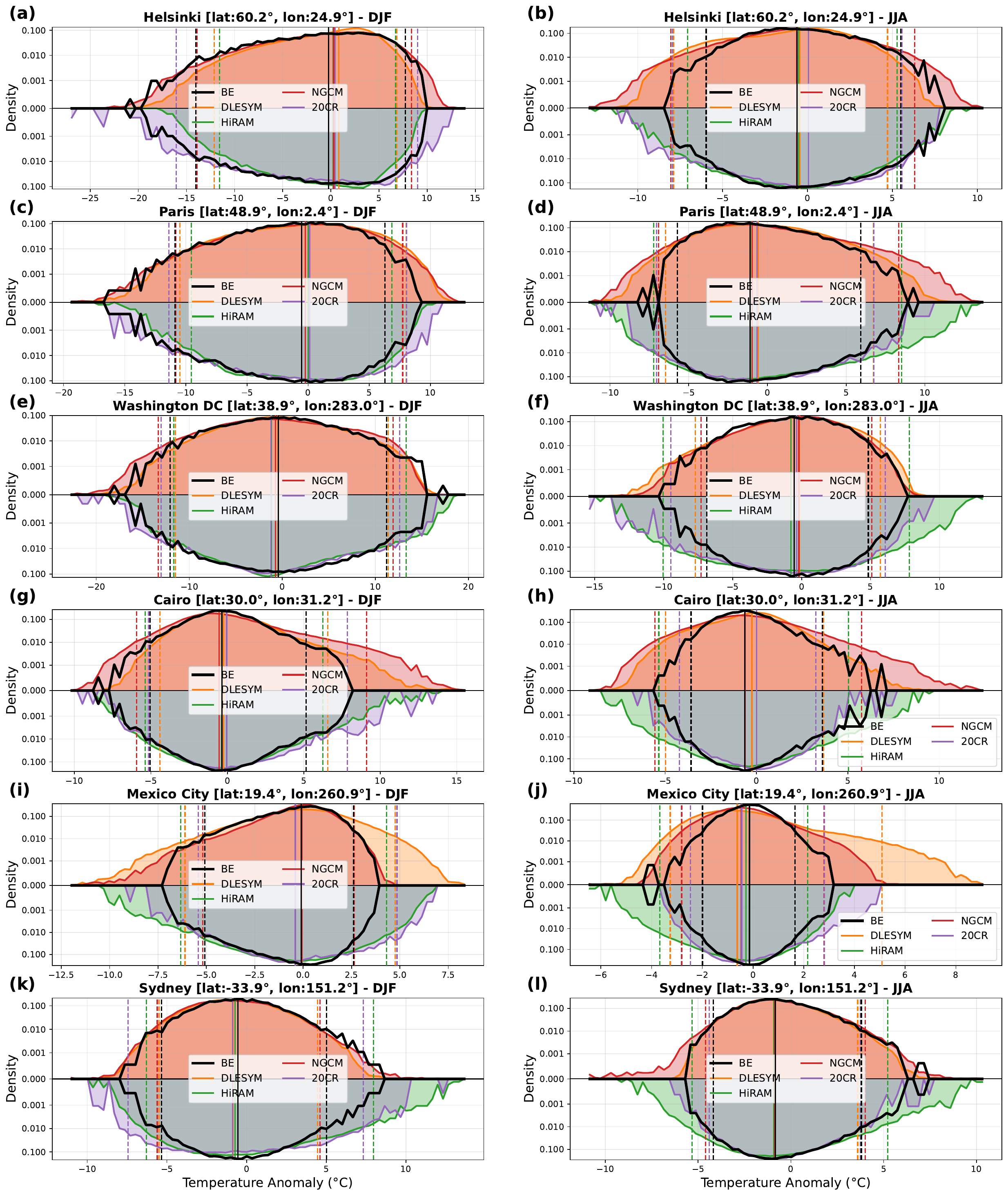}
    \caption{\textbf{Distribution of temperature anomalies from 1900--1960.} 
    Log-scaled temperature probability density distributions for DJF (left) and JJA (right) at Helsinki, Paris, Washington DC, Cairo, Mexico City, and Sydney. Anomalies are computed relative to the 1980--2010 reference period. Black denotes Berkeley Earth (BE), red denotes NGCM, orange denotes DL\textit{ESy}M, purple denotes 20CR, and green denotes HiRAM; lines are distributed above and below the abscissa for clarity. Shading represents the distribution, with the thin vertical solid line indicating the median (50th percentile) and dashed lines indicating the 1st and 99th percentiles.}
    \label{fig:tempdist}
\end{figure}

Our verification analysis is based on the heatwave and coldwave definitions from \citet{hirsch2021cmip6} and \citet{perkinsIncreasingFrequencyIntensity2012}, which depend  on the 90th and 10th percentile temperature thresholds, respectively. To complement these reanalyses-based percentile verification statistics, we now compare model simulations  directly with heatwave statistics for surface temperature observations from Berkeley Earth (BE) \citep{rohde2020berkeley} at six capital cities---Helsinki, Paris, Washington DC, Cairo, Mexico City, and Sydney---chosen because BE daily data covers 99.9\% of the out-of-sample period (1900--1960). Figure~\ref{fig:tempdist} shows DJF and JJA anomaly distributions (relative to 1980--2010). Despite not being trained on early-20th-century time period, the DL models reproduce the cold tails well. In particular, NGCM closely matches the DJF cold extremes in Helsinki, Paris, Washington DC, and Sydney, from 99th percentile to 99.9th percentile, even outperforming HiRAM and 20CR. The heatwave tail for these cities also shows relatively good agreement with the BE results. In contrast, for the tropical cities (Cairo and Mexico City), the DL models exhibit broader distributions and excessive probability density in the distribution tails, which may relate to higher persistence/autocorrelation in the tropics, suggesting a source of regional discrepancy that warrants further analysis.

\section{Conclusion and Discussion} \label{sec:conclusion}

We have evaluated the ability of deep-learning (DL) general circulation models (NGCM and DL\textit{ESy}M) to simulate land heatwaves and coldwaves during 1900--2010, comparing their performance with a physical model (HiRAM) using simulations following an AMIP protocol. Despite being trained on ERA5 data from 1980--2020, both DL models successfully reproduce the frequency and spatial patterns of heatwave and coldwave events during the out-of-sample period (1900--1960), with skill comparable to HiRAM, and mostly high correlations with reanalysis verification (20CRv3). Exceptions include the tropics, where we find that the models have higher autocorrelation that 20CRv3, and North Asia and central North America, which we speculate may be due to changes in land-surface conditions not present in any of the models. Furthermore, by directly comparing temperature distributions across six cities, we show that the DL models are able to capture the tails of the temperature distributions (temperature extremes).

The degree of physical components appears to affect temperature extremes frequency through the autocorrelation of the surface temperature field. The fully data-driven model DL\textit{ESy}M tends to overestimate extreme event frequency and has the highest temporal autocorrelation, while the physics–DL hybrid NGCM produces event frequencies more aligned with HiRAM, and correspondingly similar autocorrelation. The higher temperature autocorrelation in fully DL-based models (DL\textit{ESy}M and ACE2-ERA5) may result from the absence of explicit physical constraints and a tendency to produce smoother, more persistent temperature anomalies due to learning objectives—such as loss functions (e.g., mean squared error)—that favor minimizing short-term variability. Additionally, the large ensemble sizes enabled by the efficiency of DL models improve the stability and robustness of their climatological estimates, particularly compared to HiRAM’s limited 5-member ensemble.

These results demonstrate that DL-based models can offer credible simulations of temperature extremes. Their computational efficiency also enables large ensembles for better uncertainty quantification. However, limitations remain in terms of differing temporal autocorrelation from physical models, and lack of predictive skill to changes in land surface conditions and radiative forcing. Hybrid architectures and expanded training datasets may help improve upon these limitations. Overall, deep learning models represent a promising and complementary approach to traditional GCMs in climate modeling, particularly for simulating and analyzing climate extremes.

\section*{Open Research Section}

The plotting package and EOF calculation SACPY \citep{meng2021research,meng2023sacpy,meng2025coupled} is located at \url{https://github.com/ZiluM/sacpy}. The NGCM is located at \url{https://neuralgcm.readthedocs.io/en/latest/} and the DL\textit{ESy}M is located at \url{https://github.com/AtmosSci-DLESM/DLESyM}. The 20CRv3 dataset is located at \url{https://www.esrl.noaa.gov/psd/data/20thC_Rean/}. The HadISST dataset is located at \url{https://www.metoffice.gov.uk/hadobs/hadisst/}. The ERA5 is located at \url{https://cds.climate.copernicus.eu/cdsapp\#/home}.

\section*{Acknowledgments}
This research was supported by NSF awards 2402475, 2202526 and 2105805, and Heising-Simons Foundation award 2023-4715. We acknowledge high-performance computing support from the Casper cluster (\url{https://doi.org/10.5065/qx9a-pg09}) provided by NCAR's Computational and Information Systems Laboratory, sponsored by the National Science Foundation. The Copernicus Climate Data Store provided access to ERA5. We acknowledge the conversations with Nathaniel Cresswell-Clay (University of Washington), Dale Durran (University of Washington), Oliver Watt-Meyer (AI2 company), Dmitrii Kochkov (Google Research), Vince Cooper (University of Washington), and Dominik Stiller (University of Washington) on the model configuration and AMIP experiments running. Z.M. also acknowledges the conversations with Zhanxiang Hua (University of Washington) and Satveer Sandhu (University of Washington) about the heatwave and coldwave index calculation. 

\newpage

%
%

\bibliography{ref}

\begin{thebibliography}{53}
\providecommand{\natexlab}[1]{#1}
\providecommand{\url}[1]{\texttt{#1}}
\renewcommand{\UrlFont}{\rmfamily}
\providecommand{\urlprefix}{URL }
\expandafter\ifx\csname urlstyle\endcsname\relax
  \providecommand{\doi}[1]{https://doi.org/\discretionary{}{}{}#1}\else
  \providecommand{\doi}{https://doi.org/\discretionary{}{}{}\begingroup \urlstyle{rm}\Url}\fi
\providecommand{\eprint}[2][]{\url{#2}}

\bibitem[{Bao et~al.(2020)Bao, Zhang, Zhao,, and Chen}]{baoImpactObservationData2020}
Bao, X., F.~Zhang, Y.~Zhao, and Y.~Chen, 2020: The impact of the observation data assimilation on atmospheric reanalyses over tibetan plateau and western yunnan-guizhou plateau. \textit{Atmosphere}, \textbf{12~(1)}, 38, \doi{10.3390/atmos12010038}.

\bibitem[{Bi et~al.(2023)Bi, Xie, Zhang, Chen, Gu,, and Tian}]{biAccurateMediumrangeGlobal2023}
Bi, K., L.~Xie, H.~Zhang, X.~Chen, X.~Gu, and Q.~Tian, 2023: Accurate medium-range global weather forecasting with 3d neural networks. \textit{Nature}, \textbf{619~(7970)}, 533--538, \doi{10.1038/s41586-023-06185-3}.

\bibitem[{Canton(2021)}]{canton2021world}
Canton, H., 2021: World meteorological organization—wmo. \textit{The Europa Directory of International Organizations 2021}, Routledge, 388--393.

\bibitem[{Chan and Huybers(2019)Chan, and Huybers}]{chanSystematicDifferencesBucket2019}
Chan, D., and P.~Huybers, 2019: Systematic differences in bucket sea surface temperature measurements among nations identified using a linear-mixed-effect method. \textit{Journal of Climate}, \textbf{32~(9)}, 2569--2589, \doi{10.1175/JCLI-D-18-0562.1}.

\bibitem[{Chan et~al.(2021)Chan, Vecchi, Yang,, and Huybers}]{Chan2021}
Chan, D., G.~A. Vecchi, W.~Yang, and P.~Huybers, 2021: Improved simulation of 19th- and 20th-century north atlantic hurricane frequency after correcting historical sea surface temperatures. \textit{Science Advances}, \textbf{7~(26)}, eabg6931, \doi{10.1126/sciadv.abg6931}, \eprint{https://www.science.org/doi/pdf/10.1126/sciadv.abg6931}.

\bibitem[{Chen et~al.(2023)Chen, Zhong, Zhang, Cheng, Xu, Qi,, and Li}]{chen2023fuxi}
Chen, L., X.~Zhong, F.~Zhang, Y.~Cheng, Y.~Xu, Y.~Qi, and H.~Li, 2023: Fuxi: A cascade machine learning forecasting system for 15-day global weather forecast. \textit{npj climate and atmospheric science}, \textbf{6~(1)}, 190.

\bibitem[{Chien et~al.(2025)Chien, Barnes,, and Maloney}]{chienModulationTropicalCyclogenesis2025}
Chien, M.-T., E.~Barnes, and E.~Maloney, 2025: Modulation of tropical cyclogenesis on subseasonal-to-interannual timescales in the deep-learning climate emulator ace2. \doi{10.31223/X5NF15}.

\bibitem[{{Cresswell-Clay} et~al.(2024){Cresswell-Clay}, Liu, Durran, Liu, Espinosa, Moreno,, and Karlbauer}]{cresswell-clayDeepLearningEarth2024}
{Cresswell-Clay}, N., B.~Liu, D.~Durran, A.~Liu, Z.~I. Espinosa, R.~Moreno, and M.~Karlbauer, 2024: arXiv:2409.16247. A deep learning earth system model for stable and efficient simulation of the current climate. arXiv, \doi{10.48550/arXiv.2409.16247}, \eprint{2409.16247}.

\bibitem[{Domeisen et~al.(2022)}]{domeisenPredictionProjectionHeatwaves2022}
Domeisen, D. I.~V., and Coauthors, 2022: Prediction and projection of heatwaves. \textit{Nature Reviews Earth \& Environment}, \textbf{4~(1)}, 36--50, \doi{10.1038/s43017-022-00371-z}.

\bibitem[{Donat et~al.(2016)Donat, King, Overpeck, Alexander, Durre,, and Karoly}]{donatExtraordinaryHeat1930s2016}
Donat, M.~G., A.~D. King, J.~T. Overpeck, L.~V. Alexander, I.~Durre, and D.~J. Karoly, 2016: Extraordinary heat during the 1930s us dust bowl and associated large-scale conditions. \textit{Climate Dynamics}, \textbf{46~(1-2)}, 413--426, \doi{10.1007/s00382-015-2590-5}.

\bibitem[{Eyring et~al.(2016)Eyring, Bony, Meehl, Senior, Stevens, Stouffer,, and Taylor}]{eyring2016overview}
Eyring, V., S.~Bony, G.~A. Meehl, C.~A. Senior, B.~Stevens, R.~J. Stouffer, and K.~E. Taylor, 2016: Overview of the coupled model intercomparison project phase 6 (cmip6) experimental design and organization. \textit{Geoscientific Model Development}, \textbf{9~(5)}, 1937--1958.

\bibitem[{Gates et~al.(1999)}]{gates1999overview}
Gates, W.~L., and Coauthors, 1999: An overview of the results of the atmospheric model intercomparison project (amip i). \textit{Bulletin of the American Meteorological Society}, \textbf{80~(1)}, 29--56.

\bibitem[{Gessner et~al.(2021)Gessner, Fischer, Beyerle,, and Knutti}]{gessner2021very}
Gessner, C., E.~M. Fischer, U.~Beyerle, and R.~Knutti, 2021: Very rare heat extremes: quantifying and understanding using ensemble reinitialization. \textit{Journal of Climate}, \textbf{34~(16)}, 6619--6634.

\bibitem[{Goodfellow et~al.(2016)Goodfellow, Bengio,, and Courville}]{goodfellow2016deep}
Goodfellow, I., Y.~Bengio, and A.~Courville, 2016: \textit{Deep Learning}. MIT press.

\bibitem[{Harris et~al.(2016)Harris, Lin,, and Tu}]{harris2016high}
Harris, L.~M., S.-J. Lin, and C.~Tu, 2016: High-resolution climate simulations using gfdl hiram with a stretched global grid. \textit{Journal of Climate}, \textbf{29~(11)}, 4293--4314.

\bibitem[{Hersbach et~al.(2020)}]{hersbachERA5GlobalReanalysis2020}
Hersbach, H., and Coauthors, 2020: The era5 global reanalysis. \textit{Quarterly Journal of the Royal Meteorological Society}, \textbf{146~(730)}, 1999--2049, \doi{10.1002/qj.3803}.

\bibitem[{Hirsch et~al.(2021)Hirsch, Ridder, Perkins-Kirkpatrick,, and Ukkola}]{hirsch2021cmip6}
Hirsch, A.~L., N.~N. Ridder, S.~E. Perkins-Kirkpatrick, and A.~Ukkola, 2021: Cmip6 multimodel evaluation of present-day heatwave attributes. \textit{Geophysical Research Letters}, \textbf{48~(22)}, e2021GL095\,161.

\bibitem[{Hua and Anderson-Frey(2023)Hua, and Anderson-Frey}]{hua2023tornadic}
Hua, Z., and A.~Anderson-Frey, 2023: How are tornadic supercell soundings significantly different from nearby baseline environments? \textit{Geophysical Research Letters}, \textbf{50~(8)}, e2022GL102\,580.

\bibitem[{Hua et~al.(2025)Hua, Hakim,, and Anderson-Frey}]{hua2025performance}
Hua, Z., G.~Hakim, and A.~Anderson-Frey, 2025: Performance of the pangu-weather deep learning model in forecasting tornadic environments. \textit{Geophysical Research Letters}, \textbf{52~(7)}, e2024GL109\,611.

\bibitem[{Jacobson et~al.(2020)Jacobson, Yang, Vecchi,, and Horowitz}]{Jacobson2020}
Jacobson, T.~W., W.~Yang, G.~A. Vecchi, and L.~W. Horowitz, 2020: Impact of volcanic aerosol hemispheric symmetry on sahel rainfall. \textit{Climate Dynamics}, \textbf{55~(7)}, 1733--1758, \doi{10.1007/s00382-020-05347-7}.

\bibitem[{Jia et~al.(2016)}]{jia2016roles}
Jia, L., and Coauthors, 2016: The roles of radiative forcing, sea surface temperatures, and atmospheric and land initial conditions in us summer warming episodes. \textit{Journal of Climate}, \textbf{29~(11)}, 4121--4135.

\bibitem[{Kochkov et~al.(2024{\natexlab{a}})}]{kochkovNeuralGeneralCirculation2024a}
Kochkov, D., and Coauthors, 2024{\natexlab{a}}: Neural general circulation models for weather and climate. \textit{Nature}, \textbf{632~(8027)}, 1060--1066, \doi{10.1038/s41586-024-07744-y}.

\bibitem[{Kochkov et~al.(2024{\natexlab{b}})}]{neuralgcm2024}
Kochkov, D., and Coauthors, 2024{\natexlab{b}}: Neural general circulation models for weather and climate. \urlprefix\url{https://neuralgcm.readthedocs.io/en/latest/index.html}, accessed: 2025-04-28.

\bibitem[{Lam et~al.(2023)}]{lam2023learning}
Lam, R., and Coauthors, 2023: Learning skillful medium-range global weather forecasting. \textit{Science}, \textbf{382~(6677)}, 1416--1421.

\bibitem[{Legg(2021)}]{legg2021ipcc}
Legg, S., 2021: Ipcc, 2021: Climate change 2021-the physical science basis. \textit{Interaction}, \textbf{49~(4)}, 44--45.

\bibitem[{Meng et~al.(2025)Meng, Hakim,, and Steig}]{meng2025coupled}
Meng, Z., G.~J. Hakim, and E.~J. Steig, 2025: Coupled seasonal data assimilation of sea ice, ocean, and atmospheric dynamics over the last millennium. \textit{Journal of Climate}.

\bibitem[{Meng et~al.(2021)Meng, Hu, Ai, Zhang,, and Shan}]{meng2021research}
Meng, Z., Z.~Hu, Z.~Ai, Y.~Zhang, and K.~Shan, 2021: {Research on Planar Double Compound Pendulum Based on RK-8 Algorithm}. \textit{Journal on Big Data}, \textbf{3~(1)}, 11.

\bibitem[{Meng and Li(2024)Meng, and Li}]{meng2024pacific}
Meng, Z., and T.~Li, 2024: {Why is the Pacific meridional mode most pronounced in boreal spring?} \textit{Climate Dynamics}, \textbf{62~(1)}, 459--471.

\bibitem[{Meng et~al.(2023)Meng, Zhu,, and Hakim}]{meng2023sacpy}
Meng, Z., F.~Zhu, and G.~J. Hakim, 2023: {Sacpy--A Python Package for Statistical Analysis of Climate}. \textit{AGU23}.

\bibitem[{NCEI(2025)}]{ncei_billion_dollar_2025}
NCEI, N., 2025: U.s. billion-dollar weather and climate disasters (2025). \urlprefix\url{https://www.ncei.noaa.gov/access/billions/}, accessed: 2025-04-28, \doi{10.25921/stkw-7w73}.

\bibitem[{O'Neill et~al.(2016)}]{o2016scenario}
O'Neill, B.~C., and Coauthors, 2016: The scenario model intercomparison project (scenariomip) for cmip6. \textit{Geoscientific Model Development}, \textbf{9~(9)}, 3461--3482.

\bibitem[{Orlowsky and Seneviratne(2012)Orlowsky, and Seneviratne}]{orlowsky2012global}
Orlowsky, B., and S.~I. Seneviratne, 2012: Global changes in extreme events: regional and seasonal dimension. \textit{Climatic change}, \textbf{110}, 669--696.

\bibitem[{Perkins et~al.(2012)Perkins, Alexander,, and Nairn}]{perkinsIncreasingFrequencyIntensity2012}
Perkins, S.~E., L.~V. Alexander, and J.~R. Nairn, 2012: Increasing frequency, intensity and duration of observed global heatwaves and warm spells. \textit{Geophysical Research Letters}, \textbf{39~(20)}, \doi{10.1029/2012GL053361}.

\bibitem[{Poli et~al.(2016)}]{poliERA20CAtmosphericReanalysis2016}
Poli, P., and Coauthors, 2016: Era-20c: An atmospheric reanalysis of the twentieth century. \textit{Journal of Climate}, \textbf{29~(11)}, 4083--4097, \doi{10.1175/JCLI-D-15-0556.1}.

\bibitem[{Ragone et~al.(2018)Ragone, Wouters,, and Bouchet}]{ragone2018computation}
Ragone, F., J.~Wouters, and F.~Bouchet, 2018: Computation of extreme heat waves in climate models using a large deviation algorithm. \textit{Proceedings of the National Academy of Sciences}, \textbf{115~(1)}, 24--29.

\bibitem[{Rayner et~al.(2003)Rayner, Parker, Horton, Folland, Alexander, Rowell, Kent,, and Kaplan}]{rayner2003global}
Rayner, N.~A., D.~E. Parker, E.~Horton, C.~K. Folland, L.~V. Alexander, D.~Rowell, E.~C. Kent, and A.~Kaplan, 2003: Global analyses of sea surface temperature, sea ice, and night marine air temperature since the late nineteenth century. \textit{Journal of Geophysical Research: Atmospheres}, \textbf{108~(D14)}.

\bibitem[{Reichstein et~al.(2019)Reichstein, {Camps-Valls}, Stevens, Jung, Denzler, Carvalhais,, and {Prabhat}}]{reichsteinDeepLearningProcess2019a}
Reichstein, M., G.~{Camps-Valls}, B.~Stevens, M.~Jung, J.~Denzler, N.~Carvalhais, and {Prabhat}, 2019: Deep learning and process understanding for data-driven earth system science. \textit{Nature}, \textbf{566~(7743)}, 195--204, \doi{10.1038/s41586-019-0912-1}.

\bibitem[{Rohde and Hausfather(2020)Rohde, and Hausfather}]{rohde2020berkeley}
Rohde, R.~A., and Z.~Hausfather, 2020: The berkeley earth land/ocean temperature record. \textit{Earth System Science Data Discussions}, \textbf{2020}, 1--16.

\bibitem[{Ronneberger et~al.(2015)Ronneberger, Fischer,, and Brox}]{ronneberger2015u}
Ronneberger, O., P.~Fischer, and T.~Brox, 2015: U-net: Convolutional networks for biomedical image segmentation. \textit{Medical image computing and computer-assisted intervention--MICCAI 2015: 18th international conference, Munich, Germany, October 5-9, 2015, proceedings, part III 18}, Springer, 234--241.

\bibitem[{Rossow and Duenas(2004)Rossow, and Duenas}]{rossow2004international}
Rossow, W., and E.~Duenas, 2004: The international satellite cloud climatology project (isccp) web site: An online resource for research. \textit{Bulletin of the American Meteorological Society}, \textbf{85~(2)}, 167--172.

\bibitem[{Seneviratne et~al.(2021)}]{seneviratne2021weather}
Seneviratne, S.~I., and Coauthors, 2021: Weather and climate extreme events in a changing climate.

\bibitem[{Sippel et~al.(2024)}]{sippelEarlytwentiethcenturyColdBias2024}
Sippel, S., and Coauthors, 2024: Early-twentieth-century cold bias in ocean surface temperature observations. \textit{Nature}, \textbf{635~(8039)}, 618--624, \doi{10.1038/s41586-024-08230-1}.

\bibitem[{Slivinski et~al.(2021)}]{slivinski2021evaluation}
Slivinski, L.~C., and Coauthors, 2021: An evaluation of the performance of the twentieth century reanalysis version 3. \textit{Journal of Climate}, \textbf{34~(4)}, 1417--1438.

\bibitem[{Taylor et~al.(2012)Taylor, Stouffer,, and Meehl}]{taylor2012overview}
Taylor, K.~E., R.~J. Stouffer, and G.~A. Meehl, 2012: An overview of cmip5 and the experiment design. \textit{Bulletin of the American meteorological Society}, \textbf{93~(4)}, 485--498.

\bibitem[{Tebaldi et~al.(2006)Tebaldi, Hayhoe, Arblaster,, and Meehl}]{tebaldi2006going}
Tebaldi, C., K.~Hayhoe, J.~M. Arblaster, and G.~A. Meehl, 2006: Going to the extremes: an intercomparison of model-simulated historical and future changes in extreme events. \textit{Climatic change}, \textbf{79}, 185--211.

\bibitem[{Ullrich et~al.(2024)}]{ullrich2024recommendations}
Ullrich, P.~A., and Coauthors, 2024: Recommendations for comprehensive and independent evaluation of machine learning-based earth system models. \textit{arXiv preprint arXiv:2410.19882}.

\bibitem[{Van~Loon et~al.(2025)Van~Loon, Rugenstein,, and Barnes}]{van2025reanalysis}
Van~Loon, S., M.~Rugenstein, and E.~A. Barnes, 2025: Reanalysis-based global radiative response to sea surface temperature patterns: Evaluating the ai2 climate emulator. \textit{arXiv preprint arXiv:2502.10893}.

\bibitem[{Vonich and Hakim(2024)Vonich, and Hakim}]{vonich2024predictability}
Vonich, P.~T., and G.~J. Hakim, 2024: Predictability limit of the 2021 pacific northwest heatwave from deep-learning sensitivity analysis. \textit{Geophysical Research Letters}, \textbf{51~(19)}, e2024GL110\,651.

\bibitem[{{Watt-Meyer} et~al.(2024)}]{watt-meyerACE2AccuratelyLearning2024}
{Watt-Meyer}, O., and Coauthors, 2024: arXiv:2411.11268. Ace2: Accurately learning subseasonal to decadal atmospheric variability and forced responses. arXiv, \doi{10.48550/arXiv.2411.11268}, \eprint{2411.11268}.

\bibitem[{Wu and Xue(2024)Wu, and Xue}]{wuDataDrivenWeatherForecasting2024}
Wu, Y., and W.~Xue, 2024: Data-driven weather forecasting and climate modeling from the perspective of development. \textit{Atmosphere}, \textbf{15~(6)}, 689, \doi{10.3390/atmos15060689}.

\bibitem[{Yang et~al.(2021)Yang, Hsieh,, and Vecchi}]{Yang2021}
Yang, W., T.-L. Hsieh, and G.~A. Vecchi, 2021: Hurricane annual cycle controlled by both seeds and genesis probability. \textit{Proceedings of the National Academy of Sciences}, \textbf{118~(41)}, e2108397\,118, \doi{10.1073/pnas.2108397118}.

\bibitem[{Yang et~al.(2019)Yang, Vecchi, Fueglistaler, Horowitz, Luet, Mu{\~n}oz, Paynter,, and Underwood}]{Yang2019}
Yang, W., G.~A. Vecchi, S.~Fueglistaler, L.~W. Horowitz, D.~J. Luet, A.~G. Mu{\~n}oz, D.~Paynter, and S.~Underwood, 2019: Climate impacts from large volcanic eruptions in a high-resolution climate model: The importance of forcing structure. \textit{Geophysical Research Letters}, \textbf{46~(13)}, 7690--7699, \doi{https://doi.org/10.1029/2019GL082367}.

\bibitem[{Zhao et~al.(2009)Zhao, Held, Lin,, and Vecchi}]{zhaoSimulationsGlobalHurricane2009}
Zhao, M., I.~M. Held, S.-J. Lin, and G.~A. Vecchi, 2009: Simulations of global hurricane climatology, interannual variability, and response to global warming using a 50-km resolution gcm. \textit{Journal of Climate}, \textbf{22~(24)}, 6653--6678, \doi{10.1175/2009JCLI3049.1}.

\end{thebibliography}
\bibliographystyle{ametsocV6}

%
%
%
%
%

\section*{Supporting Information (SI)}

\raggedbottom
\onecolumn
\pagestyle{plain}
%
%
%
%
%
%
%





%
%


\title{Supporting Information for ``Evaluation of Out-of-Sample Land Heat and Cold Waves in Deep Learning-Based Atmospheric Models"}

\noindent\textbf{Contents of this file}
\begin{enumerate}
\item Figures S1 to S12
\item Table S1 
\end{enumerate}
\setcounter{figure}{0}
\renewcommand{\thefigure}{S\arabic{figure}}

\begin{table}[h!]
\centering
\caption{Summary of Models and Experimental Setup.}
\begin{threeparttable}

\begin{tabularx}{\linewidth}{@{}lXXX@{}}
\toprule
\textbf{Attribute} & \textbf{NGCM} & \textbf{DL\textit{ESy}M} & \textbf{HiRAM} \\
\midrule
\textbf{Model Type}         & Physics–DL Hybrid  & Purely Data-Driven & Physical GCM \\
\textbf{Horizontal Resolution} &  $\sim$ 2.8°                       & $\sim$ 1°                             & $\sim$50 km (C180) \\
\textbf{Boundary Forcing}    & SST \& SIC                  & SST                        & SST \& SIC \\
\textbf{Training Period}     & 1979–2017 (ERA5)             & 1983–2016 (ERA5/ISCCP)            & N/A (Physics-based) \\
\textbf{Ensemble Size}       & 100                          & 100                               & 5 \\
\bottomrule
\end{tabularx}
\end{threeparttable}
\label{tab1}
\end{table}

\begin{figure}
    \centering
    \includegraphics[width=\linewidth]{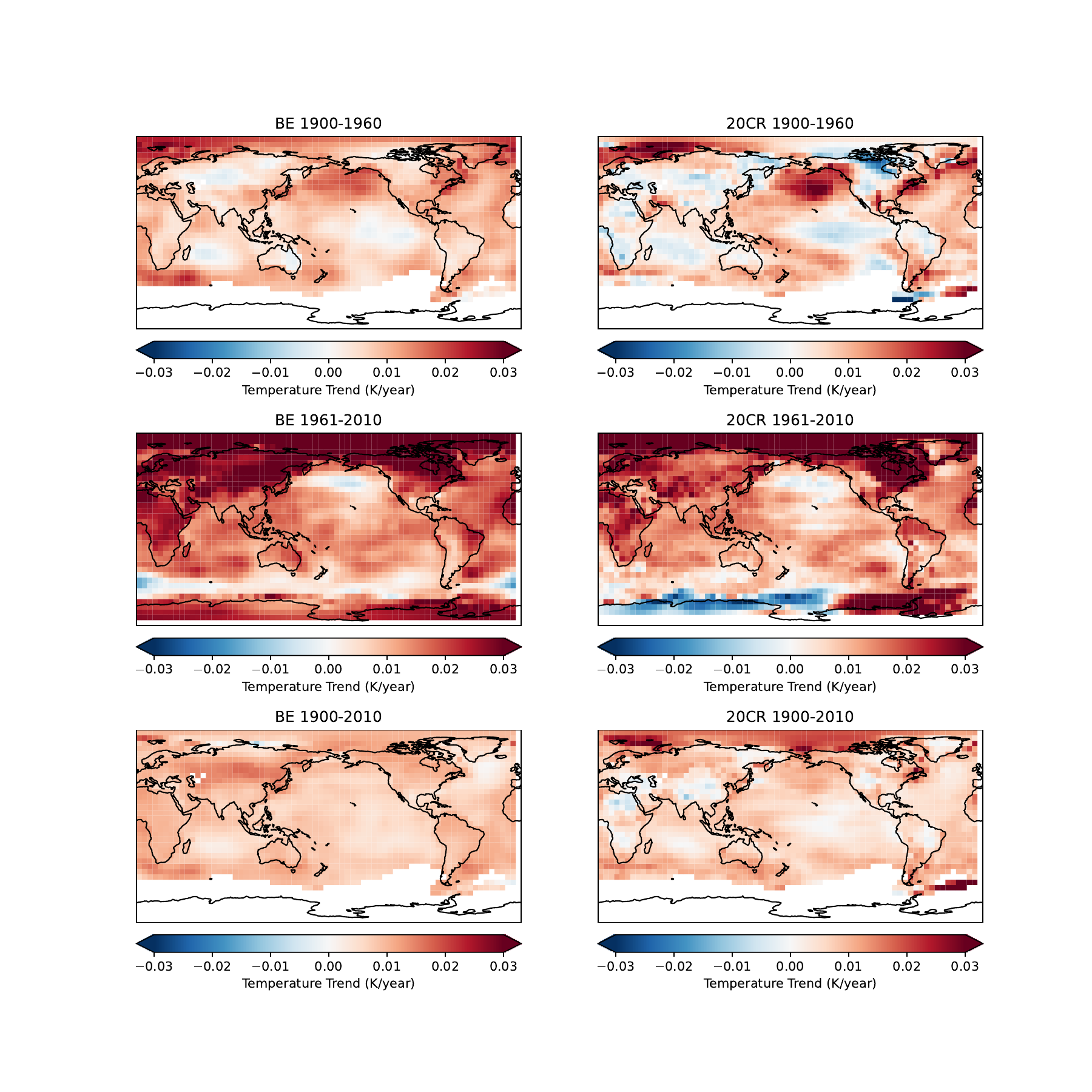}
    \caption{Annual-mean temperature trends from 20CR (left) and Berkeley Earth (BE, right) for the periods 1900--1960, 1961--2010 and 1900--2010.}
    \label{sfig:temp_trend}
\end{figure}

\begin{figure}
    \centering
    \includegraphics[width=\linewidth]{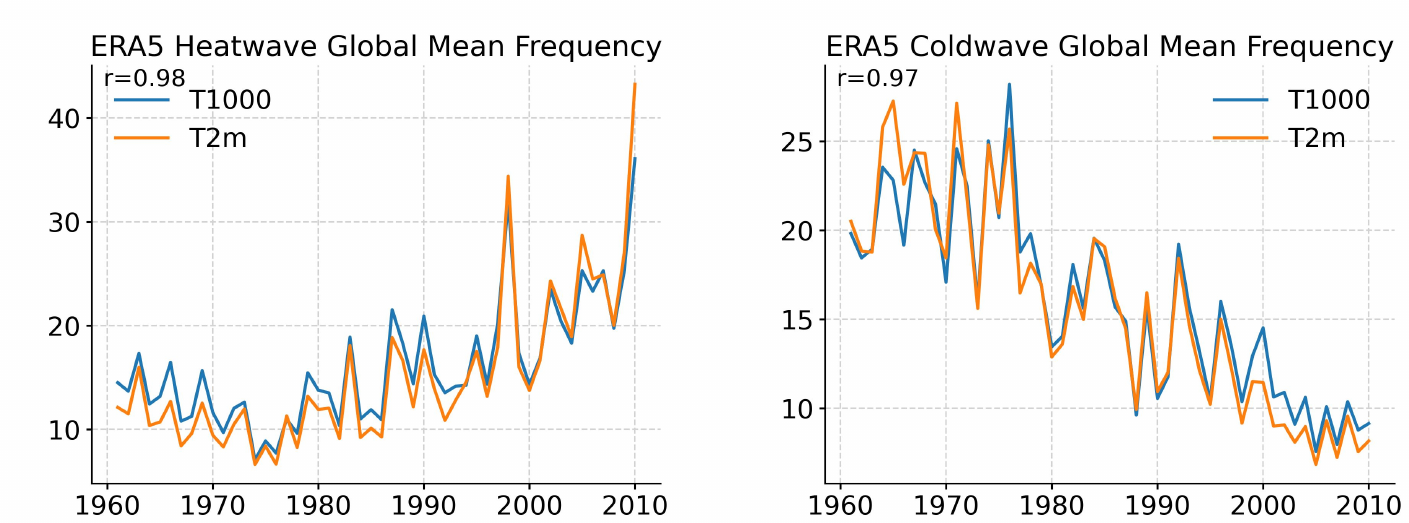}
    \caption{Annual mean heatwave (coldwave) frequency as a function of time from 1961 to 2010 from 1000hPa temperature calculation (blue) and 2-meter air temperature calculation (yellow).}
    \label{sfig:era5t1000t2}
\end{figure}

\begin{figure}
    \centering
    \includegraphics[width=\linewidth]{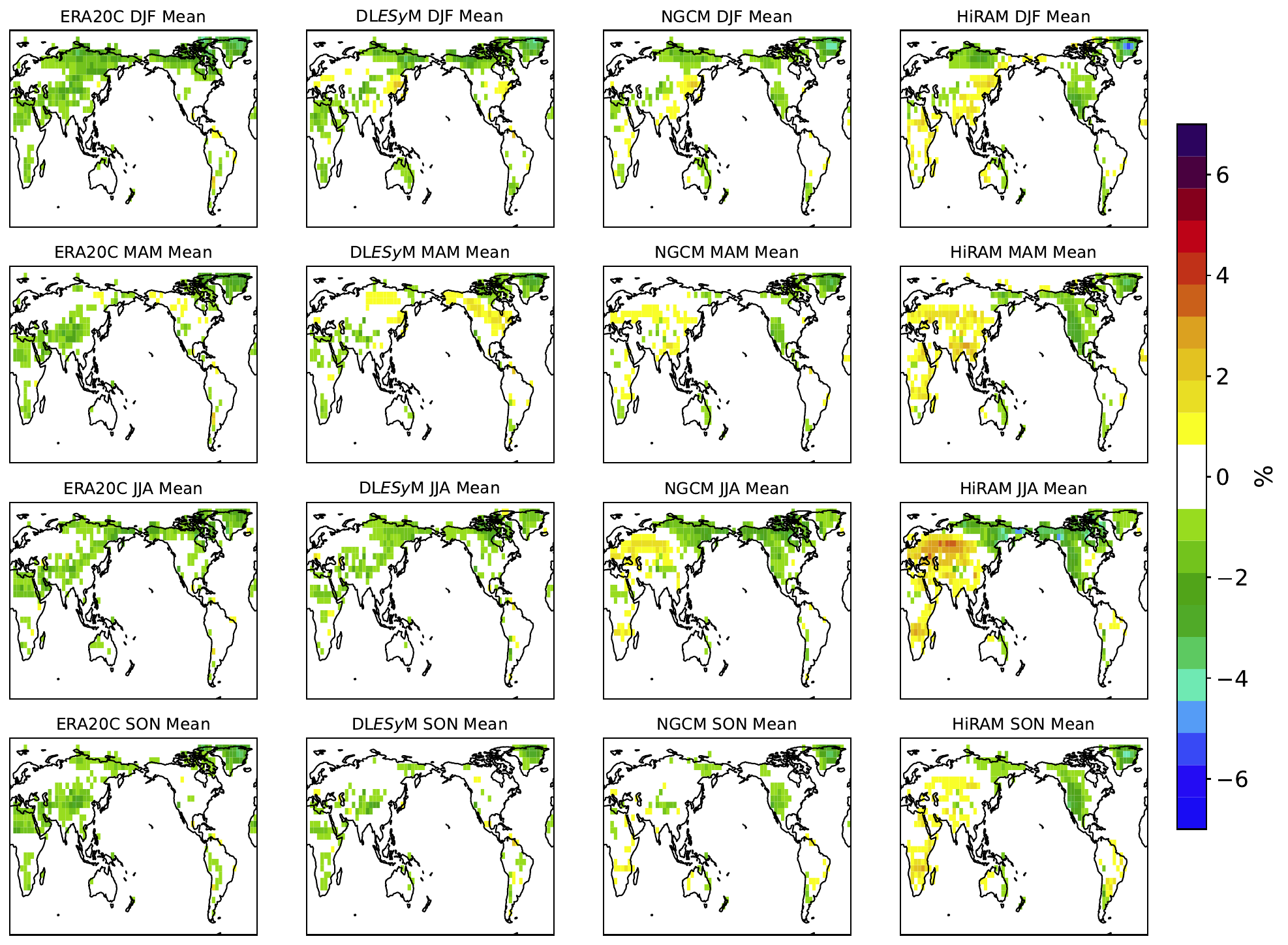}
    \caption{The ratio of heatwave threshold (90\%) differences between ERA20C, DLESYM, NGCM, and HiRAM compared to 20CR, calculated as (x - 20CR) / 20CR in DJF, MAM, JJA and SON.}
    \label{sfig:hwthres}
\end{figure}

\begin{figure}
    \centering
    \includegraphics[width=\linewidth]{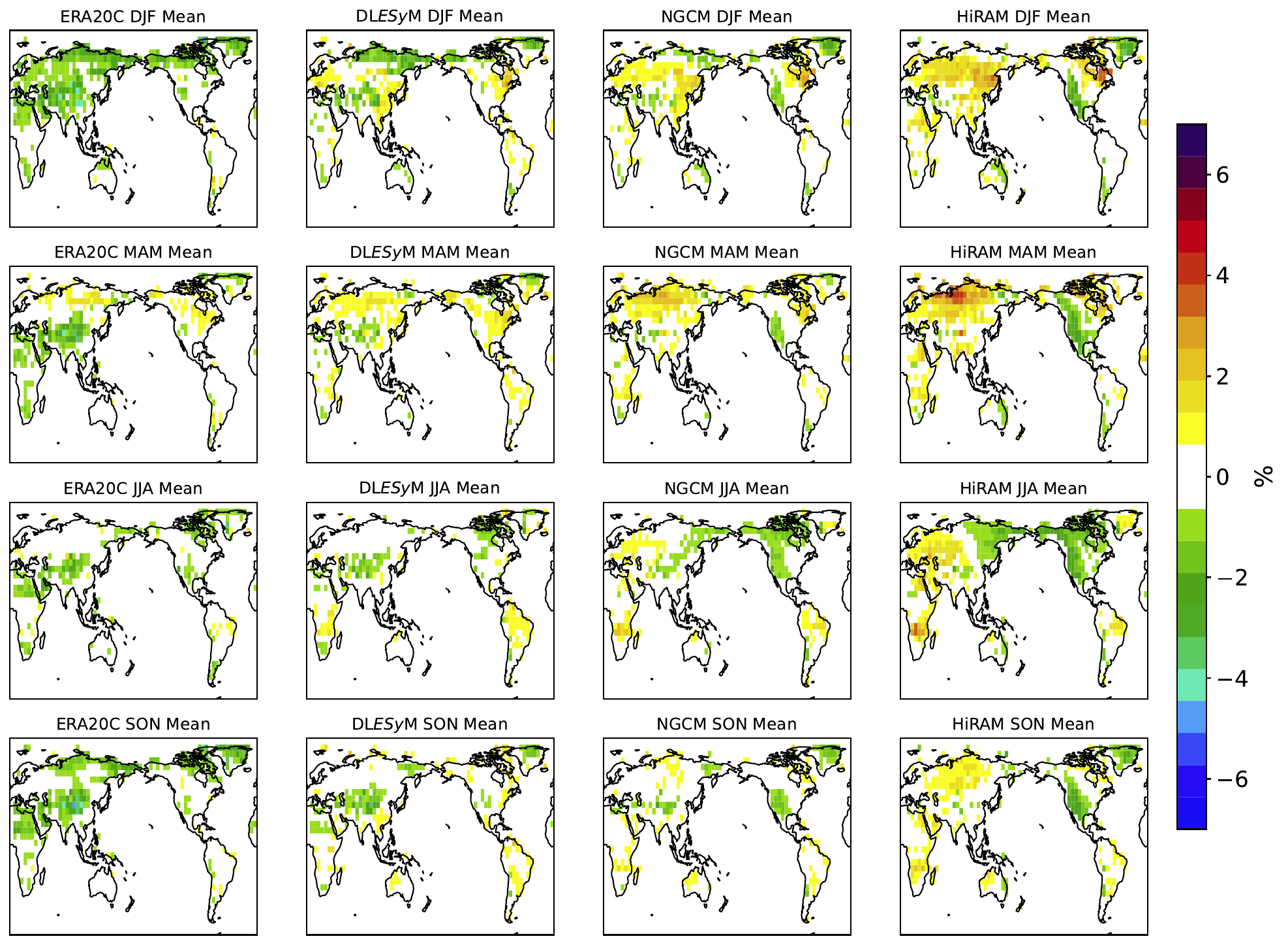}
    \caption{Same as Figure~\ref{sfig:hwthres}, but for the coldwave threshold (10\%) .}
    \label{sfig:cwthres}
\end{figure}

\begin{figure}
    \centering
    \includegraphics[width=1\linewidth]{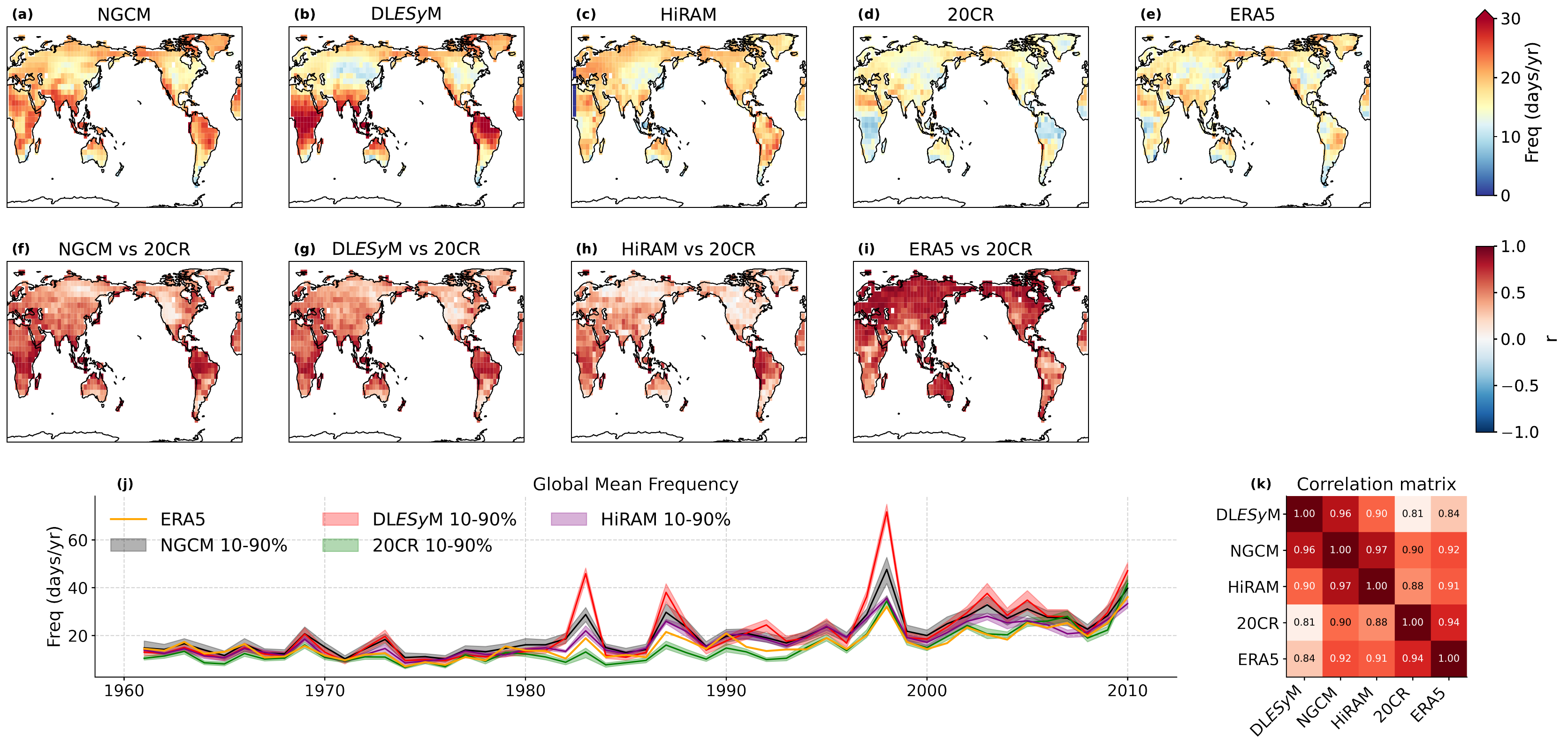}
    \caption{(a--e) Time-average heatwave frequency from 1961 to 2010 in NGCM, DL\textit{ESy}M, HiRAM, 20CR, and ERA5. (f--i) Correlation of annual mean heatwave frequency with 20CRv3. (j) Annual mean heatwave frequency as a function of time from 1961 to 2010. (k) Correlation matrix of global mean heatwave frequency between each model, 20CRv3, and ERA5.}
    \label{sfig:currentHW}
\end{figure}
\begin{figure}
    \centering
    \includegraphics[width=1\linewidth]{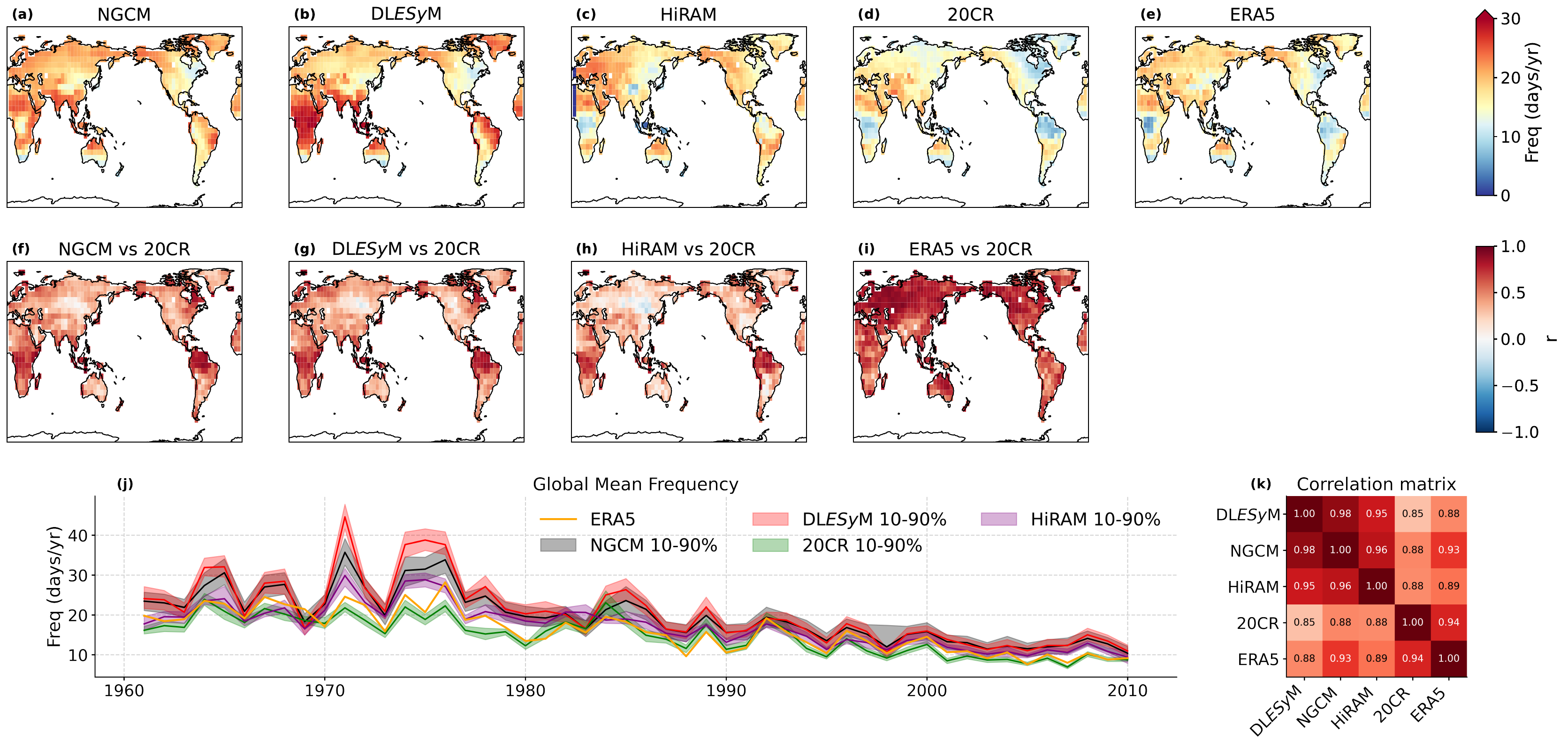}
    \caption{Same as Figure \ref{sfig:currentHW} but for coldwaves.}
    \label{sfig:currentCW}
\end{figure}

\begin{figure}
    \centering
    \includegraphics[width=1\linewidth]{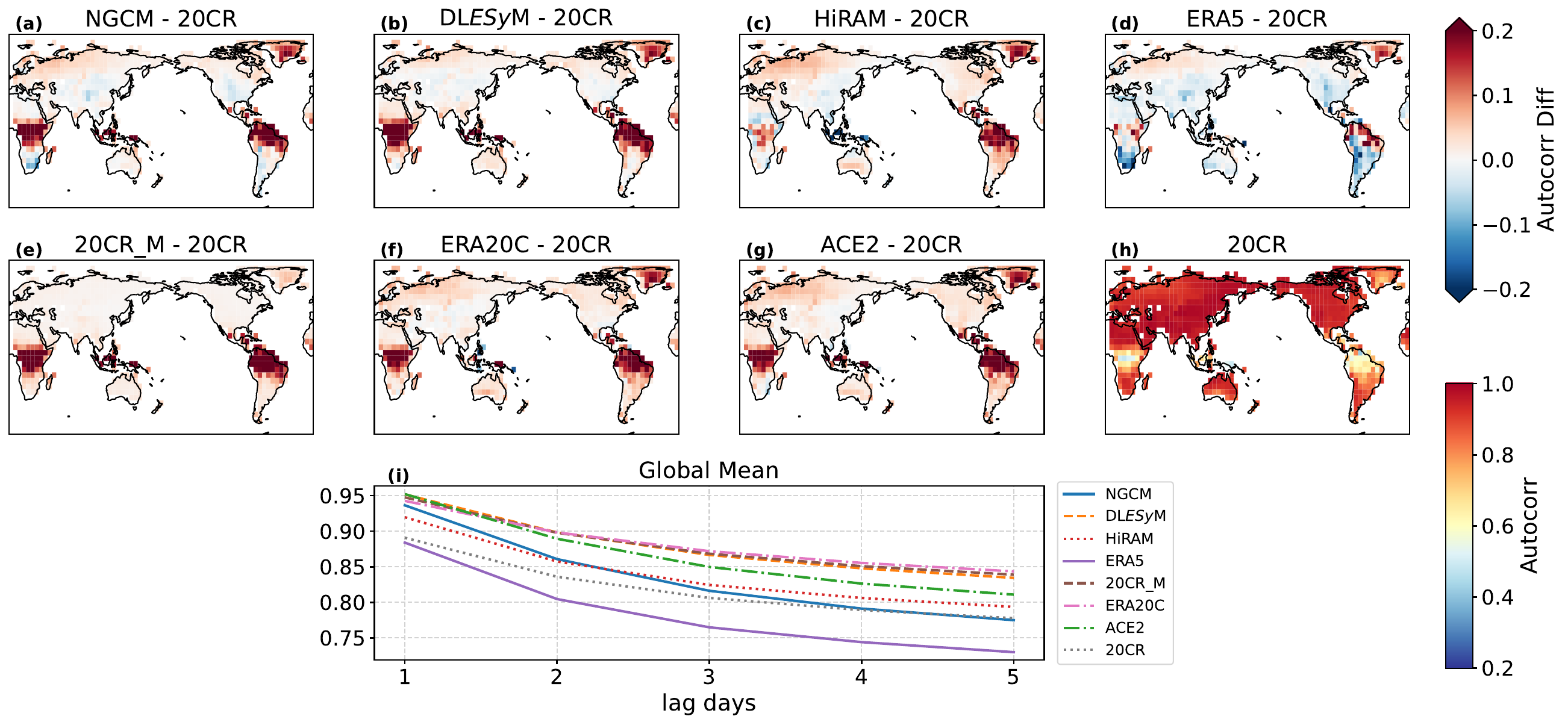}
    \caption{(a--g) 1-day lag autocorrelation difference with 20CRv3. (h) 1-day lag autocorrelation of 20CRv3. (i) Global mean land temperature autocorrelation as a function of time.}
    \label{sfig:currentAuto}
\end{figure}

\begin{figure}
    \centering
    \includegraphics[width=0.5\linewidth]{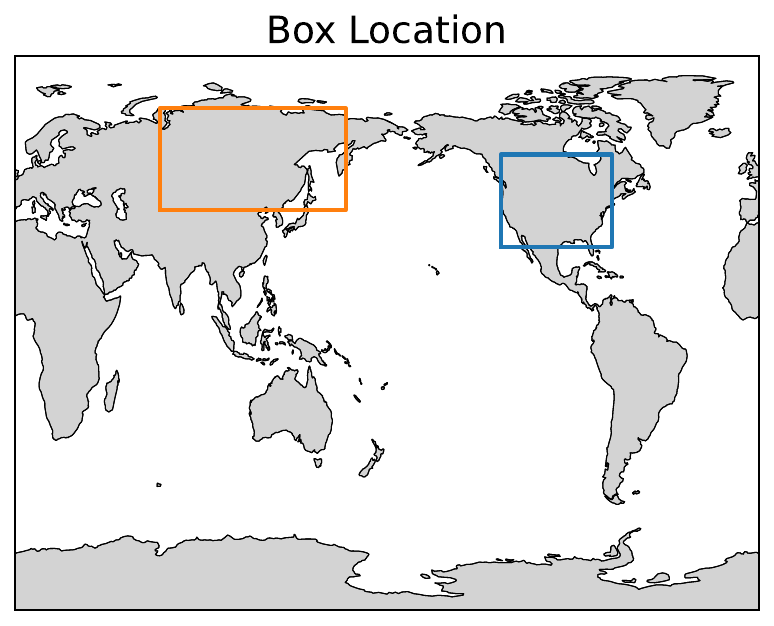}
    \caption{American Region (orange) and North Asian Region (blue).}
    \label{sfig:region}
\end{figure}
\begin{figure}
    \centering
    \includegraphics[width=\linewidth]{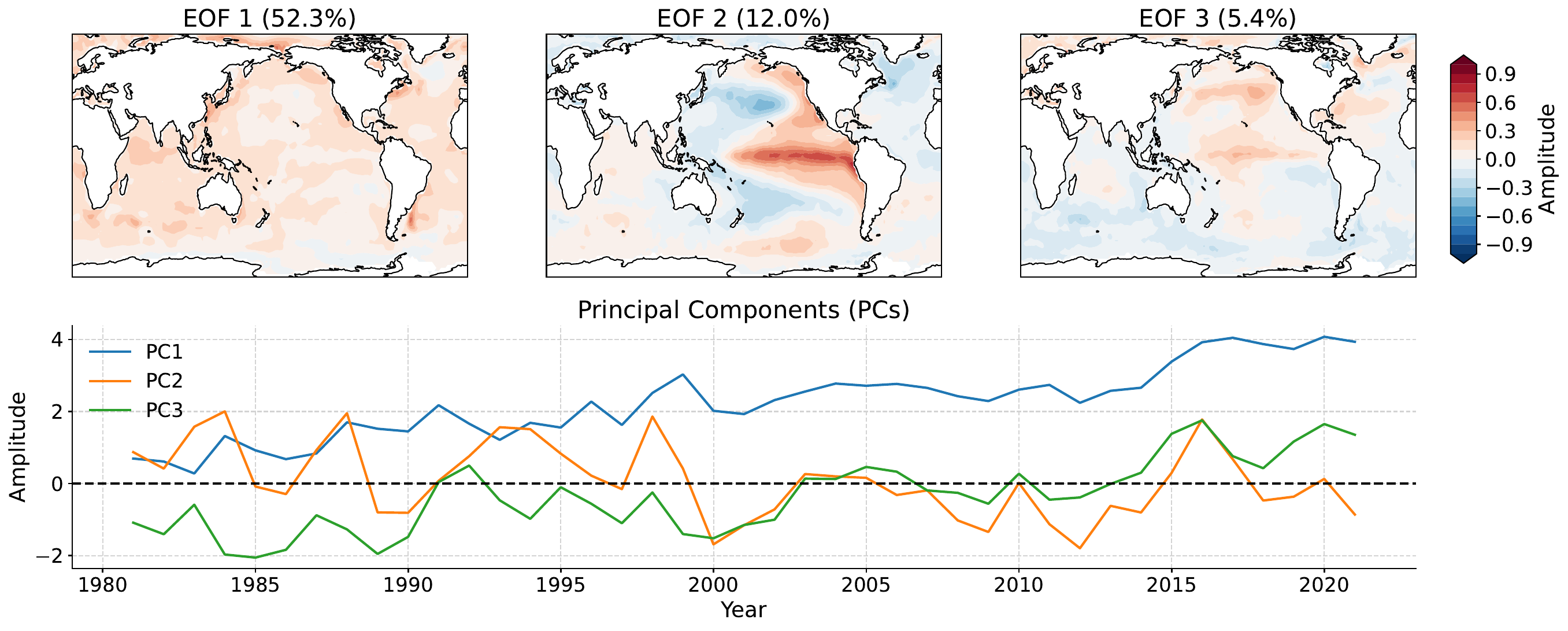}
    \caption{\textbf{EOF Patterns and Principal Components.} The first three Empirical Orthogonal Function (EOF) modes (upper panel) and their corresponding principal components (PCs; lower panel).}
    \label{sfig:eof}
\end{figure}
\begin{figure}
    \centering
    \includegraphics[width=0.5\linewidth]{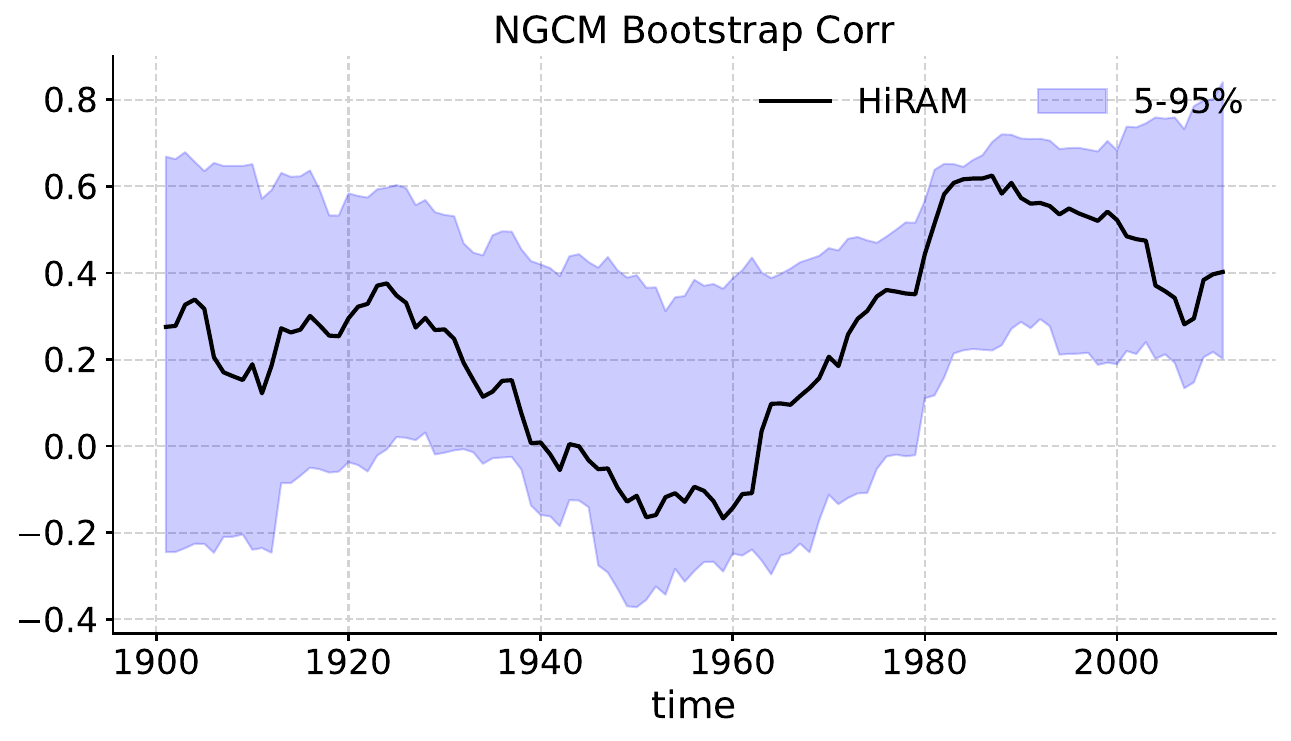}
    \caption{Bootstrap sampling of five-member NGCM simulations to approximate the distribution of five-member HiRAM results. The black line denotes the 40-year rolling correlation of American temperature between HiRAM and Berkeley Earth. The blue shading indicates the 5th to 95th percentile range of the same rolling correlation derived from the NGCM bootstrap distribution.}
    \label{sfig:bstest}
\end{figure}

\begin{figure}
    \centering
    \includegraphics[width=0.7\linewidth]{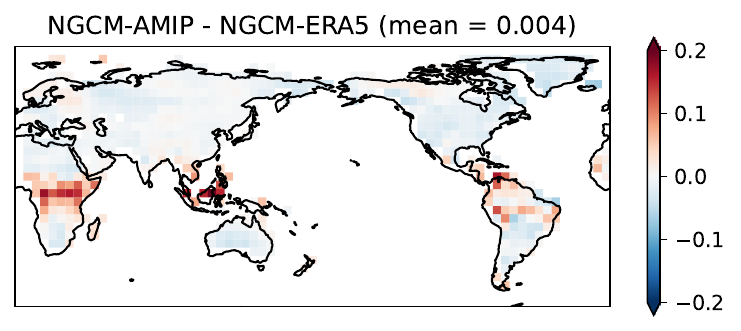}
    \caption{The 1-day lag autocorrelation difference between those using daily SST obtained through interpolation of monthly HadISST data and NGCM AMIP experiments forced with daily SST from ERA5 from 2010 to 2020.}
    \label{sfig:ERA5-AMIP}
\end{figure}

\begin{figure}
    \centering
    \includegraphics[width=1\linewidth]{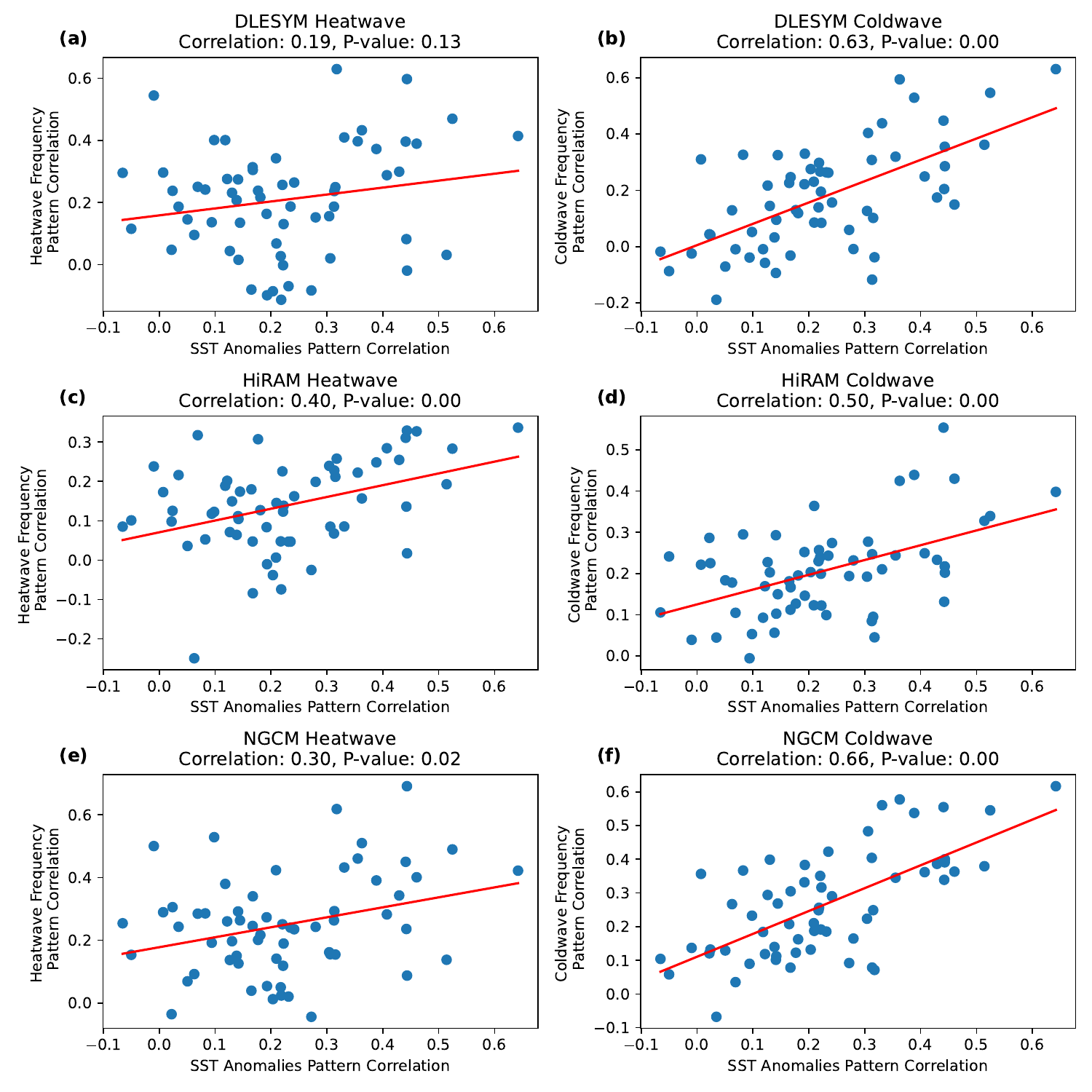}
    \caption{\textbf{Link between SST analogue similarity and extremes frequency similarity.}  
    For each out-of-sample year (1900–1960), the in-sample year (1979–2017) with the highest SST anomaly pattern correlation is identified as its analogue year. Scatter plots show the relationship between SST anomaly similarity (correlation, x-axis) and the model-simulated frequency pattern similarity of heatwaves (left panels) and coldwaves (right panels) for DLESYM, HiRAM, and NGCM. 
    Red lines indicate linear regression fits, with panel titles reporting correlation coefficients and $p$-values. Panel labels (a–f) correspond to each model–extreme combination.}
    \label{sfig:corrsst}
\end{figure}

\end{document}